\documentclass[a4paper,fleqn]{cas-sc}

\usepackage[authoryear]{natbib}
\def\tsc#1{\csdef{#1}{\textsc{\lowercase{#1}}\xspace}}
\tsc{WGM}
\tsc{QE}
\begin{document}
\let\WriteBookmarks\relax
\def\floatpagepagefraction{1}
\def\textpagefraction{.001}

\shorttitle{Evaluation of SOTA Deep Learning Architectures for Aerodynamical Predictions}    

\shortauthors{}  

\title [mode = title]{Evaluation of State-of-the-Art Deep Learning Architectures for Aerodynamical Predictions}  




\author{Jan Scherz}%[<options>]

\cormark[1]

\fnmark[1]

\ead{Jan.Scherz@dlr.de}


\credit{Conceptualization, Methodology, Software, Validation, Formal analysis, Investigation, Data Curation, Writing - Original Draft, Writing - Review \& Editing, Visualization}

\author{Derrick Hines}%[]

% Footnote of the second author
\fnmark[2]

% Email id of the second author
\ead{Derrick.HinesChaves@dlr.de}


\credit{Validation, Resources, Data Curation, Writing - Review \& Editing}

\affiliation{organization={DLR German Aerospace Center},
            addressline={Lilienthalpl. 7}, 
            city={Brunswick},
%           citysep={}, % Uncomment if no comma needed between city and postcode
            postcode={38108}, 
            state={Lower Saxony},
            country={Germany}}

\author{Philipp Bekemeyer}%[]

% Footnote of the second author
\fnmark[3]

% Email id of the second author
\ead{Philipp.Bekemeyer@dlr.de}


\credit{Conceptualization, Methodology, Validation, Resources, Investigation, Data Curation, Writing - Review \& Editing, Supervision, Project administration, Funding acquisition}

\cortext[1]{Corresponding author}

\fntext[1]{Research Scientist, Institute of Aerodynamics and Flow Technology, Center for Computer Applications in Aerospace Science and Engineering}
\fntext[2]{Team Leader, Institute of Aerodynamics and Flow Technology, Center for Computer Applications in Aerospace Science and Engineering}
\fntext[3]{Research Scientist, Institute of Aerodynamics and Flow Technology, Center for Computer Applications in Aerospace Science and Engineering}


\begin{abstract}
Surrogate models are used to substitute classical numerical solvers in engineering applications where the computational cost of the latter becomes infeasible. For instance, in aerodynamics such models offer cost-effective alternatives to computational fluid dynamics in problems such as shape optimization and load analysis, which oftentimes require high-fidelity simulations for a multitude of different parameter combinations. A specific class of deep learning-based surrogate models termed operator learning models directly approximates the solution operators to the partial differential equations underlying the physical phenomenon, thereby learning to replicate solutions to entire families of problems. However, while nowadays numerous architectures of this type get published, corresponding benchmark studies remain scarce. In this article, we advance the study of artificial intelligence-based surrogate methods by thoroughly benchmarking four state-of-the-art operator learning models on their aptitude for applications in aerospace engineering. In two experiments, we assess the models' capabilities of predicting the surface pressure distribution on two-dimensional airfoil shapes of varying complexity and on an industrial-scale three-dimensional aircraft configuration. Thereby, we evaluate the models' abilities to fulfill frequent requirements in aerodynamics such as capturing discontinuities (shocks) in the solutions, scalability towards excessive amounts of mesh points and handling of data scarcity. Accompanied by a careful analysis, our findings drive forward the field of artificial intelligence-based surrogate modeling by providing detailed insights into the strengths and weaknesses of the individual architectures, thus allowing to identify priorities for future developments. In particular the Bi-Stride Multi-Scale Graph Neural Network and Transolver(++) are highlighted as promising surrogate models for aerodynamical applications. \nocite{*}%% Remove this line from your manuscript.
\end{abstract}

% Keywords

\maketitle

\section{Introduction}

A wide range of complex physical phenomena can be modeled mathematically by means of \textit{partial differential equations} (PDEs) \cite{evans}. In aerodynamics, the solving of problems often relies on simulations using numerical methods to approximate the solution to the PDEs underlying the physical process of interest \cite{numerics, roubicek}. Such simulations, however, tend to be computationally expensive. For example, the design optimization of aircraft wings typically involves computational fluid dynamics simulations for various shape parameters. However, performing several hundreds or even thousands of simulations, where each of them takes several hours or days, is simply infeasible. This issue is tackled by \textit{surrogate models}, which constitute simplified mathematical models capable of approximating high-fidelity simulations at lower computational cost. A particular approach to surrogate modeling is given by deep learning architectures, known for strong generalization properties: After being trained on data representing the problem for a subset of the potential parameter choices, such models promise to predict solutions for any combination of parameters in the design space at high accuracy.

One popular example of deep learning architectures employed for approximating solutions to PDEs is given by \textit{Graph Neural Networks} (GNN) \cite{gnn1, gnn2}. GNNs comprehend discretized physical domains as graphs of nodes containing physical information, which is passed as messages between the nodes along connecting edges. In this way, they are able to capture local interactions between different domain points, which can be crucial for understanding the behavior of the solution to be learned. Advanced models, such as the \textit{Bi-Stride Multi-Scale Graph Neural Network} (BSMS-GNN) \cite{bsmsgnn,derrickbsmsgnn}, make use of intricate multi grid schemes, exchanging information across varying distances, ensuring the learning of global relationships without overstraining the computational capacities.

In the past years, the focus has shifted from the approximation of solutions to individual PDEs to the field of \textit{operator learning}, in which models are constructed to directly approximate the associated solution operators. Operator learning models offer the possibility to not only learn single solutions, but solutions to entire classes of problems parameterized e.g.\ by different boundary or initial conditions. In particular, a class of architectures termed \textit{neural operators} has gained traction \cite{noarticle}. Neural operators constitute a special case of operator learning models with the property of \textit{discretization invariance}, allowing them to generalize to finer discretizations than the ones they were trained on. Originally, neural operators were introduced as architectures based on integral kernel operators with learnable kernels in the form of \textit{Graph Neural Operators} (GNO) \cite{mgno, gno}. GNOs can be understood as special cases of classical GNNs. More specifically, by truncating the integral domain in a suitable way, the integral operators become equivalent to a message passing step in a GNN. The truncated domain - typically chosen as a ball of a specified radius - determines the neighborhood of (discrete) points from which a considered domain point receives messages in this process. The special feature compared to general GNNs lies in the fact that the neighborhoods - and, by this, the graphs - are constructed based on a continuous-level geometric pattern, enabling the operator to make accurate predictions on arbitrarily fine discretizations without the need for retraining. In this respect, neural operators also differ from \textit{Deep Operator Neural Networks} (DeepONet) \cite{deeponet}, another archtitecture for operator learning, which, while being able to predict solutions at arbitrary domain points, is restricted to input data on a fixed discretization

In terms of predictive accuracy, basic GNOs were soon surpassed by \textit{Fourier Neural Operators} (FNO) \cite{fno},
which make use of convolution kernels, allowing to shift the calculation of the integral into the Fourier domain via the fast Fourier transform. In the Fourier domain, the computational cost can be reduced without significant loss of information by truncating the number of considered Fourier modes. On the downside, the use of the fast Fourier transform restricts the operational capability of FNOs to uniform grids, impeding their applicability to real-world problems. This problem is for example tackled by \textit{Geometry-Aware Fourier Neural Operators} (Geo-FNO) \cite{geofno}, coupling FNOs with domain transformations and making them applicable to arbitrary geometries. The same goal is achieved by \textit{Geometry-Informed Neural Operators} (GINO) \cite{gino}, coupling FNOs with GNOs and allowing their application even on complex industrial-scale datasets. Moreover, we mention \textit{Region Interacting
Graph Neural Operators} (RIGNO) \cite{rigno}, another class of graph based operators further improving the accuracy on complex geometries at the cost of computational efficiency.

In parallel to the evolution of graph based architectures, \textit{transformer} \cite{transformers} based neural operators arose, relying on \textit{attention} to capture long-range interactions between domain points. A first attempt at making the classical attention mechanism, the cost of which scales quadratically with respect to the number of input points, applicable to PDE learning, was made in the form of \textit{Fourier-type} and \textit{Galerkin-type attention} \cite{chooseatransformer}. Based on these attention types, \textit{Operator Transformers} (OFormer) \cite{oformer} were introduced, further supplementing them by a corresponding cross attention mechanism, allowing for query locations different from the input domain points. Similarly, \textit{General Neural Operator Transformers} (GNOT) \cite{gnot} introduced \textit{heterogeneous normalized (cross) attention}, enabling in addition the processing of multiple inputs evaluated in different locations. A different strategy is pursued by \textit{Transolver} \cite{nnarticle}, which makes classical scaled dot product (self) attention feasible by applying it to a fairly small number of (learned) groups of input points subject to similar physical states, instead of directly to the input. This approach, which is referred to as \textit{Physics attention}, was further made scalable to industrial-scale data in \textit{Transolver++} \cite{transolver++}.

Moreover, many modern architectures incorporate elements of both GNOs and transformers. Models such as \textit{Universal Physics Transformer} (UPT) \cite{upt} use GNOs to encode information from the input into a reduced set of latent points, where it is processed by several transformer blocks and finally shared with arbitrary query locations via cross attention. In a similar vein, \textit{Geometry Aware Operator Transformers} (GAOT) \cite{gaot} introduced the \textit{Multiscale Attentional Graph Neural Operator} (MAGNO) encoder, employing a sophisticated combination of GNOs and attention to efficiently extract information from the input into latent tokens for a transformer processor. Furthermore, the UPT architecture was expanded in \textit{Anchored-Branched Universal Physics Transformers} (AB-UPT) \cite{abupt} with an increased focus on the processing of the query locations and on the exchange of information between volume queries, surface queries and input data. The model introduced \textit{anchor attention}, in which keys and values are only determined from a small set of anchor points subsampled from the query locations, reducing the computational cost and enhancing the operator's applicability to large-scale datasets.

Other types of neural operators include architectures based on \textit{Convolutional Neural Networks} (CNN) such as the \textit{Convolutional Neural Operator} (CNO) \cite{cno}. \textit{Decomposable Multi-scale Iterative Neural Operators }(DoMINO) \cite{domino} constitute a CNN-based architecture designed for large-scale engineering simulations capable of acting on arbitrary point clouds. Further, \textit{POSEIDON} \cite{poseidon} represents a CNN-based foundation model that has been pretrained on vast amounts of fluid dynamics data and can be finetuned efficiently to desired downstream tasks. Finally, in the form of \textit{Physics-Informed Neural Operators} (PINO) \cite{pino}, neural operators have also been investigated as architectures directly incorporating the underlying PDE into the training procedure by using its residual as an additional loss function, thereby building a bridge to \textit{Physics-Informed Neural Networks} (PINNs) \cite{pinns}.

While the field of surrogate modeling is already rich of models falling into the category of neural operators and new promising architectures keep sprouting up, benchmark studies offering fair and comprehensive evaluations of these models - in particular with respect to applications in aircraft aerodynamics - are scarce. The foundation for appropriate benchmark studies of deep learning methods for operator learning lies in suitable datasets. In aircraft aerodynamics, a variety of such datasets has been made available during recent years. The \textit{AirfRANS} dataset \cite{airfrans} provides a compilation of simulations of subsonic airflow around different two-dimensional airfoil geometries via a conjunction of the \textit{Reynolds-Averaged Navier-Stokes} (RANS) equations and the $k$\textit{-Omega Shear Stress Transport} (SST) model. The dataset introduced in \cite[Section II.C]{rae2822} consists of various airfoil geometries, created as \textit{class shape transformations} (CST) of the 2D RAE2822 airfoil used as a baseline, around which the airflow is simulated via a conjunction of the RANS equations and the \textit{Spalart-Allmaras turbulence} model. Comprising both sub- and transonic flow regions, the dataset in particular covers shocks (i.e.\ discontinuities in the solutions to the underlying PDEs), which represent a frequent difficulty in aircraft aerodynamics. In \cite{blendnet}, the same system of equations was used to create a dataset representing the airflow around various 3D blended wing body geometries. Respective datasets for automotive aerodynamics were introduced e.g.\ in form of the \textit{DrivAerML} dataset \cite{drivaerml}, the \textit{DrivAerNet} dataset \cite{drivaernet} as well as the \textit{DriveAerNet++} dataset \cite{drivaernet++}. The latter three datasets are specifically designed to simulate industrial-scale problems with numbers of (surface) mesh points ranging in the hundred thousands or even millions, pushing models to the limits of their computational efficiency. A corresponding industrial-scale dataset for aircraft aerodynamics can be found in form of the \textit{NASA CRM} dataset in \cite[Section II.D]{rae2822}, simulating the airflow around the 3D \textit{NASA Common Research Model} (NASA CRM) aircraft configuration, represented by over $450.000$ surface points, under various operational conditions. This dataset addresses a further common difficulty in industrial scenarios: Consisting of only $149$ samples, the dataset is ideal to assess models' abilities of learning to generalize to unseen data samples in the case of scarce training data. A benchmark study comparing the capabilities of multiple machine learning based surrogate models - such as random forests, gradient boosting and classical neural networks - to predict the transonic airflow around 2D airfoils is given in \cite{elrefaie}. An evaluation of classical operator learning methods (GNO, FNO, DeepONet) as well as first transformer based architectures (Galerkin-type transformer, GNOT) on various tasks, including aerodynamical predictions, is carried out in \cite{nosurvey}. Recently, an extensive standardized study thoroughly benchmarking eleven state-of-the-art models on the DrivAerNet++ dataset was presented in \cite{carbench}. Furthermore, the publication of new models is typically accompanied by an evaluation against already established architectures, cf.\ the articles cited above. However, a systematic unbiased investigation of contemporary operator learning models applied in aircraft aerodynamics yet appears to be missing.

In the present article, we are taking a step in this direction by assessing four state-of-the-art architectures on their capabilities of learning the surface pressure in the aerodynamic flow around airfoil geometries and aircraft configurations. More specifically, we carry out two experimental studies. First, we benchmark the models on the 2D airfoil dataset \cite[Section II.C]{rae2822}, thereby not only evaluating the models' aptitude for handling shocks but also their ability to generalize to unseen geometries. In order to assess the models at their peak performance, we accompany our experiment with a hyperparameter optimization, polishing their predictive capabilities by tuning the most crucial hyperparameters for each architecture. Second, we benchmark the models on the industrial-scale NASA CRM dataset \cite[Section II.D]{rae2822}, getting insights into their applicability to highly demanding real-world problems. With these two benchmark studies, our main contribution consists of the investigation of the following three critical questions in the evaluation of surrogate models' applicability to aircraft aerodynamics: 
\begin{itemize}
\item[1.)] We evaluate the models' capabilities of correctly predicting shocks.
\item[2.)] We assess the models' capabilities of handling complex geometries.
\item[3.)] We examine the models' performance when faced with typical difficulties in industrial problems, namely, data scarcity and excessive amounts of mesh points.
\end{itemize}

The outline of the article is as follows: In Section \ref{neuraloperators}, we provide a brief overview of the idea behind operator learning as well as our models of interest. In Section \ref{archsum}, we present summaries of the architectures of the individual models. In Section \ref{rae2822exp}, we carry out the experimental study on the 2D airfoil dataset: After the hyperparameter optimization in Section \ref{hpopt}, we train ten instantiations of each model in the respective optimized settings and evaluate the results in Section \ref{results2d}. In Section \ref{nasacrm}, we perform the corresponding training and evaluation of the models on the NASA CRM dataset. Finally, in Appendix \ref{A}, we present the details of the models' set ups - besides the optimized hyperparameters - used throughout our experiments (Section \ref{setup}), the exact definition of our evaluation metrics (Section \ref{metrics}) as well as a derivation of the pooling radii used in our set ups of the GNOs in the UPT and GAOT architectures (Section \ref{radiuschoices}).

\section{Problem formulation and operator learning} \label{neuraloperators}

In general, our problems of interest can be expressed in the following way: Let $\Omega \subset \mathbb{R}^d$, $d \in \{2, 3\},$ be a bounded domain, let $U,\ U^*$ be two function spaces on $\Omega$ and let $G$ be a function space on the boundary $\partial \Omega$ of $\Omega$. Given a (partial) differential operator $D_a: U \rightarrow U^*$ parametrized by $a \in \mathbb{R}^p$, $p \in \mathbb{N}$, a function $f \in U^*$, representing for example an external forcing term and a function $g \in G$, representing a boundary condition, we search for a solution $u \in U$ to the PDE
\begin{equation}
    \begin{aligned}
    D_a(u) = f \quad \quad &\text{in } \Omega, \label{pde} \\
    u = g \quad \quad &\text{in } \partial \Omega.
    \end{aligned}
\end{equation}

We point out, that in general PDEs may be posed in dependence of an additional temporal instead of only a spatial component. For the sake of simplicity and since in the present article we focus on steady-state problems, however, we neglect this fact in \eqref{pde}. If for fixed instances of the data $(\Omega, a, f, g)$ the problem \eqref{pde} possesses a unique solution $u = u(\Omega, a, f, g)$, we can introduce the associated solution operator $\mathcal{L}$,
\begin{align}
\mathcal{L}:\ (a, \Omega, f, g) \mapsto u(a, \Omega, f, g), \nonumber
\end{align}

mapping each tupel of data to the corresponding solution to \eqref{pde}. The goal in operator learning is the construction of a deep neural network $\mathcal{L}_\theta$, parametrized by learnable parameters $\theta \in \mathbb{R}^q$, $q \in \mathbb{N}$, that approximates the operator $\mathcal{L}$,
\begin{align}
\mathcal{L}_\theta \approx \mathcal{L}. \nonumber
\end{align}

In comparison to more classical approaches approximating the solution $u$ itself, operator learning models offer a decisive advantage:
\begin{itemize}
\item After being trained once, operator learning models are capable of predicting the solution to the problem \eqref{pde} for entire classes of input data (e.g.\ for arbitrary forcing terms $f \in U^*$) instead of only individual instances.
\end{itemize}

A specific class of operator learning models is constituted by neural operators. Mathematically speaking, neural operators do not only approximate discrete instantiations of the solution operator $\mathcal{L}$, but the continuous level operator $\mathcal{L}$ itself. This property gives them a second advantage over classical approaches termed \textit{discretization invariance}: While neural operators still need to be evaluated on discretizations of the physical domain, they can be evaluated at arbitrarily fine discretizations without sacrificing accuracy. For a detailed introduction to neural operators, we refer to \cite{noarticle}. In the present article, we evaluate multiple architectures falling into the class of operator learning models on their ability to learn the surface pressure distribution, determined by the RANS equations and the Spalart-Allmaras turbulence model, on varying airfoil/ aircraft geometries and under varying operating conditions, cf.\ Sections \ref{rae2822exp} and \ref{nasacrm}. More specifically, we investigate the following models:

\begin{itemize}
\item \textbf{BSMS-GNN:} The Bi-Stride Multi-Scale Graph Neural Network \cite{bsmsgnn,derrickbsmsgnn} constitutes a graph-based operator learning model comprising multiple GNNs. Starting from a given input graph, a \textit{bi-stride selection algorithm} is applied to create a range of graphs of different coarseness. Via a U-Net structure comprising a downsampling phase from the finest to the coarsest and an upsampling phase back to the finest graph, the model enables efficient long-range exchange of information between different regions of the domain. While BSMS-GNN is applicable to operator learning tasks, strictly speaking it does not fall into the neural operator class as it relies on input graphs prepared manually by the user instead of constructing graphs from the input mesh according to a continuous-level geometric pattern.
\item \textbf{Transolver:} Transolver \cite{nnarticle} represents a transformer-like neural operator architecture, in which the classical attention mechanism is replaced by Physics attention. More specifically, instead of applying scaled dot product attention directly to individual mesh points, it is applied to groups of mesh points representing physical states learned by the model. This procedure reduces the computational cost of the attention mechanism from quadradic to linear while enabling the operator to not get lost in too much information coming from too many individual mesh points.
\item \textbf{UPT:} The Universal Physics Transformer \cite{upt} constitutes a neural operator mixing elements of both transformers and graph neural networks (GNNs). An Encoder reduces the set of input points to a small number of latent tokens. These can be handled efficiently (and, in case of time-dependent problems, propagated forward in time) by the processor. A Decoder uses the processed information to predict the solution on a set of query points which may differ from the input points.
\item \textbf{GAOT:} The Geometry Aware Operator Transformer \cite{gaot} represents a neural operator designed to handle industrial-scale datasets without having to pay an accuracy-efficiency trade-off. To this end, it introduces the \textit{Multiscale Attentional Graph Neural Operator} (MAGNO) encoder, passing information across multiple scales from the input points to a small latent grid, where it can be treated efficiently by a transformer processor.
\end{itemize}

A more detailed summary of the architectures of these models is given in Section \eqref{archsum} below. We point out that UPT represents the predecessor of the more recent AB-UPT \cite{abupt}, which introduces a specific multi-branch architecture to enhance the exchange of information between volume queries, surface queries and input data. To improve the applicability to industrial-scale datasets, AB-UPT further introduces anchor attention, which reduces the computational cost by calculating keys and values only from a proportionally small subset of the actual query locations. However, as that model does not accept additional physical information besides geometrical data as input, we deem it unsuitable for our experimental studies and thus restrict our investigation to the UPT architecture. We further remark, that in the form of Transolver++ \cite{transolver++}, Transolver also possesses a successor with an increased focus on scalability to industrial-scale data. More specifically, Transolver++ constitutes a highly parallelized version of Transolver, which further introduces an improved variant of Physics attention, capable of better capturing meaningful physical states from large amounts of mesh points. Therefore, while we rely on the original Transolver in our experiments on the small scale 2D airfoil dataset in Section \ref{rae2822exp}, we employ Transolver++ for our investigations on the large scale NASA CRM dataset in Section \ref{nasacrm}.

\section{Models} \label{archsum}

In this section, we present summaries of the architectures of the models investigated in our studies. For the full details of these architectures, we refer to the respective articles. For the sake of readability, we unify the notation used in the description of the architectures across all models. Each model receives input consisting of $n_I \in \mathbb{N}$ spatial coordinates $\{x_i\}_{i=1}^{n_I} \subset \mathbb{R}^d$, $d \in \{2, 3 \}$, as well as $n_I$ physical and/ or geometric features $\{a_i\}_{i=1}^{n_I} \subset \mathbb{R}^{c_I}$, $c_I \in \mathbb{N}$, at these coordinates. The individual inputs are concatenated as $\{x_i, a_i\}_{i=1}^{n_I} \subset \mathbb{R}^{d + c_I}$. If a model is evaluated in its input coordinates $\{x_i\}_{i=1}^{n_I}$, it produces output $\{y_i\}_{i=1}^{n_I} \subset \mathbb{R}^{c_O}$, $c_O \in \mathbb{N}$. If instead the model is evaluated at different coordinates $\{x_i^O\}_{i=1}^{n_O} \subset \mathbb{R}^d$, $n_O \in \mathbb{N}$, its output is denoted by $\{y_i\}_{i=1}^{n_O} \subset \mathbb{R}^{c_O}$.

\subsection{Bi-Stride Multi-Scale Graph Neural Network}

BSMS-GNN represents a graph-based architecture consisting of several layers in which information is passed between nodes along graphs of different coarseness. While the model originally goes back to \cite{bsmsgnn}, we here employ the modification presented in \cite{derrickbsmsgnn}, which introduced edge completion at the coarsest level.

The BSMS-GNN architecture is organised in an encode-process-decode structure. The encoder consists of an MLP, which embeds the input $\{x_i, a_i\}_{i=1}^{n_I} \subset \mathbb{R}^{d + c_I}$ as latent features $\{ v_i^0 \}_{i=1}^{n_I} \subset \mathbb{R}^{c_H}$ in a hidden dimension $c_H \in \mathbb{N}$.

The processor, which is responsible for the bulk of the work, consists of a GNN of $L+1$ scales, $L \in \mathbb{N}$, arranged in a U-Net structure of $2L+1$ layers. It receives a graph $G^0$ - prepared upfront by the user and built upon the nodes $\{\tilde{x}_i^0 \}_{i=1}^{n_I} := \{x_i \}_{i=1}^{n_I}$ - as additional input, to which it associates $\{v_i^0\}_{i=1}^{n_I}$ as node features. For each scale $l = 1,...,L$, a coarser graph $G^l$ is created from the graph $G^{l-1}$ via a bi-stride selection algorithm, selecting (in each connected component) all nodes with even/ uneven - depending on which there are more of - geodesic distance to a specified seed node in $G^{l-1}$ as nodes $\{\tilde{x}_i^l \}_{i=1}^{n^l}$, $n^l \leq n^{l-1} \leq ... \leq n^0 := n_I$, for $G^l$.

The processing begins with $L$ layers of downsampling: In each layer $l \in \{0,...,L-1\}$, each node $\tilde{x}_i^l$ at first receives messages from all nodes $\tilde{x}_j^l$ in its neighborhood $N(\tilde{x}_i^l)$ in the graph $G^l$ via MLPs $\Phi_{M_l}$ and $\Phi_{U_l}$. More specifically, the node features $v_i^l$ associated to $\tilde{x}_i^l$ are updated to $\hat{v}_i^l$ via
\begin{align}
m_i^l :=& \bigoplus_{\tilde{x}_j^l \in N(\tilde{x}_i^l)} \Phi_{M_l} \left( v_i^l, v_j^l, \Delta \tilde{x}_{ji}^l, \left| \Delta \tilde{x}_{ji}^l \right| \right), \label{aggmess} \\
\hat{v}_i^l :=& v_i^l + \Phi_{U_l} \left( v_i^l, m_i^l \right), \label{interupd}
\end{align}

where the aggregation operator $\bigoplus$ can be chosen as sum or mean and $\Delta \tilde{x}_{ji}^l$ represents the difference between the coordinates $\tilde{x}_{j}^l$ and $\tilde{x}_{i}^l$. This step is followed by another update, in which the new features $v_{i}^{l+1}$ as well as an updated position $\tilde{x}_i^{l+1}$ are calculated as weighted sums
\begin{align}
v_i^{l+1} := \sum_{\tilde{x}_j^l \in N(\tilde{x}_i^l)} w_{ji}^l \hat{v}_j^l, \quad \quad \tilde{x}_i^{l+1} := \sum_{\tilde{x}_j^l \in N(\tilde{x}_i^l)} w_{ji}^l \tilde{x}_j^l. \nonumber
\end{align}

Here, the contribution edge weights $w_{ji}^l$ are calculated, at each scale $l$, to avoid overrepresentation of nodes with larger neighborhoods.

On the coarsest level $L$, message passing is carried out following the same procedure \eqref{aggmess}--\eqref{interupd} as during the downsampling phase. In contrast to the graphs $G^0, ..., G^{L-1}$, however, the graph $G^L$ introduces edges between all its nodes, to enable the exchange of information also between different connected components of the original graph $G^0$.

In the layers $L+1,...,2L$, the processing consists of upsampling: The graphs of the first $L$ layers are passed through backwards until returning to the finest graph $G^0$. In this procedure, the graph $G^l$ in each layer $l \in \{L+1,...,2L\}$ consists of the same nodes $\{\tilde{x}_i^l\}_{i=1}^{n^l} = \{\tilde{x}_i^{2L-l}\}_{i=1}^{n^{2L-l}}$ and edges as the graph $G^{2L-l}$. In each layer $l \in \{L+1,...,2L\}$, the features $v_i^l$ of the $i$-th node $\tilde{x}_i^l$ are updated via
\begin{align}
v_i^{l} &:= \sum_{j \in N(\tilde{x}_i^l)} w_{ij}^{2L-l} \hat{v}_j^{l-1} \nonumber \\
\hat{v}_i^{l} &:= \hat{v}_i^{2L - l} + \left( v_i^l + \Phi_{U_l} \left( v_i^l, m_i^l \right) \right), \nonumber
\end{align}

where the contribution weights $w_{ij}^{2L-l}$ were calculated in the corresponding layer $2L - l$ and the aggregated message $m_i^l$ is calculated according to the formula \eqref{aggmess}.

Finally, the decoder consists of another MLP, which transforms the output $\{v_i^{2L}\}_{i=1}^{n_I}$ of the processor into the final predictions $\{ y_i \}_{i=1}^{n_I} \subset \mathbb{R}^{c_O}$. For the full details of the BSMS-GNN architecture, we refer to \cite[Section 2.4]{derrickbsmsgnn}.

\subsection{Transolver}

Transolver \cite{nnarticle} constitutes a transformer-based architecture which does not apply attention directly to input points. Instead, Transolver introduces an attention mechanism named Physics attention, in which groups of input points considered to be subject to similar physical states are summarized as \textit{slizes}. Subsequently, classical scaled dot product attention is applied to establish correlations between these slizes instead of between input points. The main purpose of Physics attention lies in a reduction of the computational cost in comparison to classical attention applied in PDE learning. Indeed, by choosing the number $m \in \mathbb{N}$ of slizes significantly smaller than the number $n \in \mathbb{N}$ of input points, the computational cost of this approach can be regarded as linear (instead of quadratic) with respect to $n$.

Like BSMS-GNN, Transolver employs an encoder, consisting of a linear lifting layer, to embed its input $\{x_i, a_i\}_{i=1}^{n_I} \subset \mathbb{R}^{d + c_I}$ into latent features $\{v_i^0\}_{i=1}^{n_I} \subset \mathbb{R}^{c_H}$ of a hidden dimension $c_H$. Subsequently, the latent features are passed through $L \in \mathbb{N}$ Transolver blocks. While, as in classical transformers, all blocks comprise layer normalization, feed forward neural networks and skip connections, their core component is constituted by Physics attention: The input $\{v_i^l\}_{i=1}^{n_I} \subset \mathbb{R}^{c_H}$ to the $(l+1)$-th block is split into $m$ slizes
\begin{align}
s_j^l :=& \left\{s_{ji}^l \right\}_{i=1}^n, \quad \quad s_{ji}^l := w_{ji}^l v_i^l, \quad \quad j = 1,..., m, \nonumber
\end{align}

where $w_{ji}^l$ is to be understood as the probability of the $i$-th mesh point belonging to the $j$-th slize $s_j^l$. More specifically,
\begin{align}
w_{ji}^l := \sigma \left( \hat{v}_i^l \right)_j, \quad \quad \sigma (\xi)_j := \frac{\exp \left( \xi_j \right)}{\sum_{k=1}^m \exp \left(\xi_k \right)} \quad \text{for } \xi = ( \xi_1, ..., \xi_m ) \in \mathbb{R}^m, \nonumber
\end{align}

where $\hat{v}_i^l$ denotes a linear projection of $v_i^l$ from $\mathbb{R}^{c_H}$ onto $\mathbb{R}^m$ and $\sigma$ represents the softmax function. From each slice $s_j^l$, a \textit{Physics-aware token}
\begin{align}
z_j^l := \frac{\sum_{i=1}^n s_{ji}^l}{\sum_{i=1}^n w_{ji}^l} \nonumber
\end{align}

is created. These tokens $\{ z_j^l \}_{j=1}^m \subset \mathbb{R}^{c_H}$ are reevaluated in the context of the $m$ different physical states via classical scaled dot product self attention \textit{Attn},
\begin{align}
\{\hat{z}_j^l\}_{j=1}^m := \textit{Attn} \left(\{z_j^l\}_{j=1}^m \right) \subset \mathbb{R}^{C_H}. \nonumber
\end{align}

The updated tokens are then desliced back into features of mesh points
\begin{align}
v_i^{l+1} := \sum_{j=1}^m w_{ji}^l \hat{z}_j^l. \nonumber
\end{align}

Finally, the output $\{v_i^L\}_{i=1}^{n_I} \subset \mathbb{R}^{c_H}$ of the last Transolver block is transformed into the desired prediction $\{y_i\}_{i=1}^{n_I} \subset \mathbb{R}^{c_O}$ via a decoder consisting of another linear layer. For more details on the Transolver architecture we refer to \cite[Section 3]{nnarticle}.

\subsection{Universal Physics Transformer} \label{upt}

UPT \cite{upt} represents a neural operator built upon elements of both transformers and graph neural operators (GNOs) with a general design geared towards time-dependent problems. Like BSMS-GNN and Transolver, UPT follows an encode-process-decode structure, however, with a greater emphasis on the encoder. In this architecture, the encoder is used to reduce the number of input points in order to keep the computational cost of the attention mechanism in check. The encoder first embeds the input $\{x_i, a_i\}_{i=1}^{n_I} \subset \mathbb{R}^{d + c_I}$ - which is understood as information given at a starting time $t \in \mathbb{R}$ - as latent features $\{v_i\}_{i=1}^{n_I} \subset \mathbb{R}^{c_H}$, $c_H \in \mathbb{N}$. To this end, it utilizes sinusoidal position embeddings for the spatial coordinates and a linear lifting layer for the physical and geometric features. Each latent feature $v_i$ is associated to the corresponding input point $x_i$ as node feature. Furthermore, the encoder subsamples a set of $n_S < n_I$ \textit{supernodes} $\{\tilde{x}_i\}_{i=1}^{n_S} \subset \{x_i\}_{i=1}^{n_I}$ from the input points $x_i$. Subsequently, it employs a GNO to pass the embedded information $\{v_i\}_{i=1}^{n_I}$ from the input points to the supernodes. For a specified radius $r>0$, each supernode $\tilde{x}_i$ receives information from all input points $x_j$ lying within an $r$-neighborhood
\begin{align}
N_r\left( \tilde{x}_i \right) := \left\lbrace x_j \in \{ x_i \}_{i=1}^{n_I}:\ \left|\tilde{x}_i - x_j\right| \leq r \right\rbrace \nonumber
\end{align}

of $\tilde{x}_i$. Namely, each supernode $\tilde{x}_i$ is provided with a corresponding node feature
\begin{align}
\hat{z}_i :=& \bigoplus_{x_j^l \in N(\tilde{x}_i^l)} \Phi \left( v_j^l \right), \nonumber
\end{align}

where $\Phi$ represents a message passing MLP and the aggregation operator $\bigoplus$ is chosen as mean. The supernode features $\{\hat{z}_i\}_{i=1}^{n_S} \subset\mathbb{R}^{c_H}$ are to be understood as tokens, which are then passed through $L \in \mathbb{N}$ transformer blocks, followed by a perceiver, further pooling them into an even smaller set of $n_{\text{latent}} < n_s$ \textit{latent tokens}
\begin{align}
\left\{z_i \right\}_{i=1}^{n_{\text{latent}}} \subset \mathbb{R}^{c_H}. \nonumber
\end{align}

The processor of UPT, which is referred to as \textit{Approximator}, is responsible for the forward in time propagation in time-dependent problems. It consists of a transformer which maps the latent tokens $z_i = z_i(t)$ at the time $t$ to latent tokens $z_i(t') = z_i(t + \Delta t)$ for a time step $\Delta t > 0$. If the problem at hand requires predictions at more than one time point, the Approximator can be applied multiple times in succession. In time-independent settings, such as in our experiments, the Approximator can still be employed with a single propagation step.

Finally, the decoder consists of another transformer, which receives additional input in the form of output positions $\{x_i^O\}_{i=1}^{n_O} \subset \mathbb{R}^{d}$, $n_O \in \mathbb{N}$. It then creates predictions $\{y_i\}_{i=1}^{n_O} \subset \mathbb{R}^{c_O}$ of the solution at the time $t'$ in these output positions via cross attention, using the latent tokens $z_i(t')$ as keys and values and the output positions as queries. For more details on the UPT architecture, we refer to \cite[Section 3]{upt}.

\subsection{Geometry Aware Operator Transformer} \label{gaot}

GAOT \cite{gaot} constitutes a neural operator which, as UPT, incorporates elements of both transformers and GNOs in its architecture. Like UPT, GAOT possesses an encode-process-decode architecture with an encoder designed to transform the input points into a smaller set of latent tokens which can be handled by a transformer processor. More specifically, GAOT introduces the MAGNO encoder, which constitutes the main focus of its architecture. MAGNO is employed to pass information across different ranges from the input point cloud $\{x_i\}_{i=1}^{n_I}$ to a reduced set of latent points $\{\tilde{x}_i\}_{i=1}^{n_S} \subset \mathbb{R}^d$, $n_S \in \mathbb{N}$, constructed as a regular grid within the spatial domain. While this set of points is referred to as \textit{latent point cloud} by the model's authors (cf.\ \cite[Section 2]{gaot}), it is directly comparable to the set of supernodes in UPT. For consistency in our notation, we will therefore adopt the terminology \textit{supernodes} for GAOT throughout the rest of this article. After embedding the input features $\{a_i\}_{i=1}^{n_I} \subset \mathbb{R}^{c_I}$ as latent features $\{v_i\}_{i=1}^{n_I} \subset \mathbb{R}^{c_H}$, $c_H \in \mathbb{N}$, associated to the input points $\{x_i\}_{i=1}^{n_I}$, MAGNO makes use of $M \in \mathbb{N}$ GNOs to distribute this information to the supernodes across $M$ scales $m \in \{1, ..., M\}$: Each scale $m$ is associated to a radius $r_m > 0$ and each supernode $\tilde{x}_i$ receives information from within its $r_m$-neighborhoods
\begin{align}
N_{r_m}\left( \tilde{x}_i \right) := \left\lbrace x_j \in \{ x_i \}_{i=1}^{n_I}:\ \left|\tilde{x}_i - x_j\right| \leq r_m \right\rbrace,\quad \quad m = 1,...,M. \nonumber
\end{align}

More specifically, each supernode $\tilde{x}_i$ is provided with $M$ (physical) node features
\begin{align}
\hat{z}^p_{m, i} := \sum_{x_j \in N_{r_m}( \tilde{x}_i )} \alpha_j^m \Phi^m \left( \tilde{x}_i, x_j, v_j \right) \Phi \left(v_j \right), \nonumber
\end{align}

where $\Phi^m$ and $\Phi$ denote message passing MLPs and the weights $\alpha_j^m$ are determined via an attention mechanism acknowledging the fact that not all neighbors may have the same relevance to $\tilde{x}_i$. Furthermore, on each scale $m$, MAGNO encodes additional statistical geometric information (such as the number of neighbors of a supernode $\tilde{x}_i$, the average distance of the neighbors to $\tilde{x}_i$, etc.) separately as \textit{novel geometry embeddings} $\{\hat{z}^g_{m, i}\}_{i=1}^{n_S}$ via MLPs. The physical and geometric features are subsequently concatenated on each scale and processed by another MLP, resulting in \textit{scale-specific latent features} $\{\hat{z}_{m, i}\}_{i=1}^{n_S}$. Ultimately, the output $\{ z_i \}_{i=1}^{n_S}$ of MAGNO in the supernode is calculated as weighted averages of the scale-specific features,
\begin{align}
z_i := \sum_{m=1}^{M} \beta_{m, i} \hat{z}_{m,i}, \nonumber
\end{align}

with weights $\beta_{m, i}$ determined via softmax normalization to account for the fact that different scales may have more or less importance for different supernodes $\tilde{x}_i$.

The output of the encoder is also referred to as \textit{geometry aware tokens} $\{ z_i^0 \}_{i=1}^{n_S} := \{ z_i \}_{i=1}^{n_S}$. In the processor, these tokens are passed through $L \in \mathbb{N}$ classical transformer blocks stacked on top of each other, to exchange information between the supernodes. Utilizing flash attention and patching for increased computational efficiency, each transformer block $l \in \{1,..., L\}$ results in updated tokens $\{ z_i^l \}_{i=1}^{n_S}$. Finally, in the decoder, a backwards implementation of MAGNO is instantiated to pass the updated information $\{ z_i^L \}_{i=1}^{n_S}$ back from the supernodes to arbitrary query points $\{x_i^O\}_{i=1}^{n_O} \subset \mathbb{R}^d$, $n_O \in \mathbb{N}$, resulting in predictions $\{y_i\}_{i=1}^{n_O} \subset \mathbb{R}^{c_O}$.

When dealing with time-dependent problems, GAOT can further receive a lead time as additional input and follow one out of several possible time stepping schemes, while its architecture remains unchanged. More details on the architecture of GAOT are to be found in \cite[Section 2]{gaot}.

\section{Surface pressure prediction on 2D airfoils} \label{rae2822exp}

In our first experiment, we investigate the models' abilities to predict the surface pressure distribution on a variety of two-dimensional airfoil geometries. More specifically, we consider the dataset \cite[Section II.C]{rae2822}, comprising $598$ airfoil geometries created by applying class shape function transformations (CST) to a baseline geometry, namely, the transonic airfoil RAE2822. Each class shape transformation consists of a - fixed - class function, determining the general geometry type (airfoil), a shape function, specifying the shape of the lower and upper surfaces of the airfoil via $9$ variable parameters, as well as a - fixed - supplementary term reflecting the trailing edge thickness. The parameters varying the airfoil shapes are chosen according to a Sobol sequence within certain bounds ensuring that the geometries do not degenerate. The geometry of each sample in the dataset is determined by the coordinates of $512$ surface points resulting from these CST parametrizations. Besides the spatial coordinates, each data sample further includes the corresponding surface normal vectors, three global parameters - the Mach number $M$, the Reynolds number $Re$ and the angle of attack $AoA$ - as well as the associated pressure coefficient $C_p$ and the skin friction coefficients $C_{f_x}$, $C_{f_z}$ (in both spatial directions) in each surface point. Here, the values for the quantities $C_p$, $C_{f_x}$ and $C_{f_z}$ were calculated from a conjunction of the RANS equations and the Spalart-Allmaras turbulence model via the CFD solver DLR TAU. Out of the $598$ data samples, $498$ fixed samples constitute the training set, while the remaining $100$ samples are to be used for testing.

In our examination, we use the airfoil geometry (spatial coordinates and surface normal vectors) as well as the global flow parameters ($M$, $Re$, $AoA$) as input features, based on which the models are trained to predict the target quantity $C_p$, cf.\ Table \ref{Tab:tab1}. The skin friction coefficients $C_{f_x}$, $C_{f_z}$ instead are neglected in our experimental set up. We apply min-max scaling to normalize all input features to the range $[0, 1]$, while leaving the output quantity unchanged.

\subsection{Hyperparameters} \label{hpopt}

In an attempt to allow each model to showcase its best performance, we carry out a hyperparameter optimization. For each architecture, we select a set of hyperparameters, the influence of which on the trained model's prediction capabilities we want to investigate. Out of the $498$ training samples, we fix $100$ randomly chosen samples for validation while keeping the remaining $398$ samples for training. Subsequently, we perform a grid search, in which each model is trained - for every associated combination of hyperparameters - on the reduced training set and evaluated on the validation set. For each architecture, the best setting of hyperparameters is established in terms of the mean absolute error (MAE) on the validation set. With this setting, ten instantiations of the corresponding model are then retrained on the entire $498$ samples of the training set and evaluated on the test set for a final assessment in Section \ref{results2d} below.

Due to high training times of multiple hours per run (cf.\ Table \ref{Tab:tabRes2deff} below) and the large amount of potentially adjustable hyperparameters coming with the complex models, we have to restrict our investigation to a subset of hyperparameters. The remaining hyperparameters are fixed in accordance with the default settings in the published code and the accompanying articles; an overview can be found in Section \ref{setup} in the appendix. We further truncated the number of training epochs during the hyperparameter optimization to $2500$ per model instance, except in the case of UPT, for which the number of training epochs was part of the investigated hyperparameters as explicated below.

In our optimization we vary two common hyperparameters for all models, namely, the maximum learning rate and the batch size used during the training procedure. The remaining model-specific parameters we take into account are discussed next:

\begin{itemize}
\item \textbf{BSMS-GNN:} BSMS-GNN constitutes the only graph based model in our study which is built upon rather classical GNNs than GNOs. In particular, it does not come with an inherent method to construct a graph from the input points following a geometric pattern. Instead, the model receives a graph which is prepared upfront by the user as input; the coarsened graphs representing the individual scales are created from this input graph via the bi-stride selection algorithm. We constructed the input graph by creating the neighborhood of any node $x$ from a specified number of nodes lying the closest to the left and right of $x$ on the surface curve. During our experiment, we varied this number of neighboring nodes as well as the model's depth represented by the number of scales. All trialed combinations of hyperparameters are listed in Table \ref{Tab:tab4}.

\item \textbf{Transolver:} Transolver is built entirely upon the (Physics-)attention mechanism, which results in a straight-forward architecture resembling closely the concatenation of classical attention blocks in transformer models. The absence of GNN-based components severely simplifies selecting hyperparameters for Transolver; in particular the need to set hyperparameters for an appropriate construction of the graphs is not needed. The hyperparameter embodying the main difference between Transolver and traditional transformers in PDE learning is given by the number of slices, into which the input points are grouped instead of being exposed to attention directly. For our investigation, we thus decided to optimize this number, the full list of hyperparameter combinations we took into consideration for Transolver is given in Table \ref{Tab:tab-1}.

\item \textbf{UPT:} For UPT we carried out two rounds of hyperparameter optimization. In the first round, besides the max.\ learning rate and the batch size, we varied the number of supernodes, to which the input nodes are subsampled, and the number of latent tokens in the perceiver block of the encoder, cf.\ Section \ref{upt}. We point out that the pooling radius, determining the neighborhoods of input nodes from which the supernodes receive their messages, is another potential candidate for optimization. However, we decided to go with the authors' of \cite{upt} suggestion of 24 neighbors per supernode on average and attempted to set the pooling radius accordingly, cf.\ Section \ref{radiuschoices} in the appendix. As the approximator in the UPT architecture plays no significant role for time-independent problems and the number of keys, values and queries in the decoder is determined by the number of latent tokens in the encoder as well as the number of query points at which the model is evaluated, we further decided to restrict the optimization to hyperparameters associated to the encoder.

After this first round of hyperparameter optimization, the results of UPT fell off in comparison to the results observed from the other models, while the evolution curve of the MAE hinted at room for further improvement. Besides the need for longer training, we attributed this to the linear decay of the learning rate in the default learning rate scheduler of the published code, cf.\ Section \ref{setup}. As a consequence, we decided to increase the number of training epochs and to use an alternate learning rate scheduler, in which the linear increase during the first ten percent of epochs is followed by exponential instead of linear decay. In a direct comparison, with the hyperparameters set in accordance with the results seen during the first round, we observed an improvement of the MAE on the validation set from originally $0.0182$ (for $2.500$ training epochs and linear decay) to $0.0159$ (for $10.000$ training epochs and linear decay) and $0.0141$ (for $10.000$ epochs and exponential decay). We then proceeded with a second round of hyperparameter optimization, in which we set the learning rate decay to exponential and varied the number of epochs, the decay rate and, once more, the maximum learning rate. The specific combinations of hyperparameters we investigated during the first and second round of hyperparameter optimization for UPT are listed in Table \ref{Tab:tab3}.

\item \textbf{GAOT:} Out of the investigated models, GAOT comes with the arguably most complex architecture, making its hyperparameter optimization the most restrictive one. In addition to the max.\ learning rate and the batch size, we considered the number of supernodes and the supernode pooling radius, determining the approximate number of neighbors per supernode, cf.\ Section \ref{gaot}. We point out that here, during the optimization of the max.\ learning rate, we further adjusted the learning rates in the individual phases of the sophisticated learning rate schedule proportionally, cf.\ Table \ref{Tab:tabHPsGAOTTrain}. The specific hyperparameter combinations we tested for GAOT are listed in Table \ref{Tab:tab2}.

We remark that the investigated numbers of supernodes for GAOT are significantly higher than the corresponding values chosen for UPT. The reason for this is as follows: In UPT, the supernodes are randomly subsampled from the input nodes and therefore lie exactly on the surface curve. In the (publicly available code of the) GAOT model, instead, the supernodes are created as a regular grid in the entire two-dimensional plane. In particular, many supernodes will be located far away from the airfoil surface. We thus set the overall number of supernodes in the GAOT model higher to ensure that sufficiently many of them will lie in the vicinity of the surface curve. We further remark that for each number of supernodes we experimented with two different pooling radii. The higher value, respectively, is chosen in an attempt to ensure - similar to our set up of UPT - at least 24 neighbors on average for the supernodes closest to the airfoil surface and to guarantee that each input node lies within the neighborhood of a supernode. Our calculation of the pooling radius in dependence of that number is outlined in Section \ref{radiuschoices} in the appendix.

Since both the encoder and the decoder of GAOT are built upon the MAGNO architecture, the same hyperparameters can be used for both of these components. Furthermore, the processor of GAOT consists of a fairly classical transformer. In our optimization we thus decided to rely on the default settings and to not take into account any further hyperparameters from these parts of the architecture. Another potential hyperparameter to be optimized is given by the number of scales in the MAGNO encoder. Experiments in this direction were restricted by high memory usage, prohibiting the use of more than two scales in an instantiation of GAOT. Instead, we retreated to the single-scale approach for our hyperparameter optimization, in accordance with the default settings. However, training a GAOT model with the best combination of hyperparameters according to the results of our opimization and an additional second scale indicated that the supplementary pooling radius only improves the outcome if it is larger than the original one.
Under these circumstances it appears more rational to employ GAOT with only the largest considered radius instead of multiple scales under higher computational cost.
\end{itemize}

\begin{figure}
\centering
\includegraphics[scale=0.38]{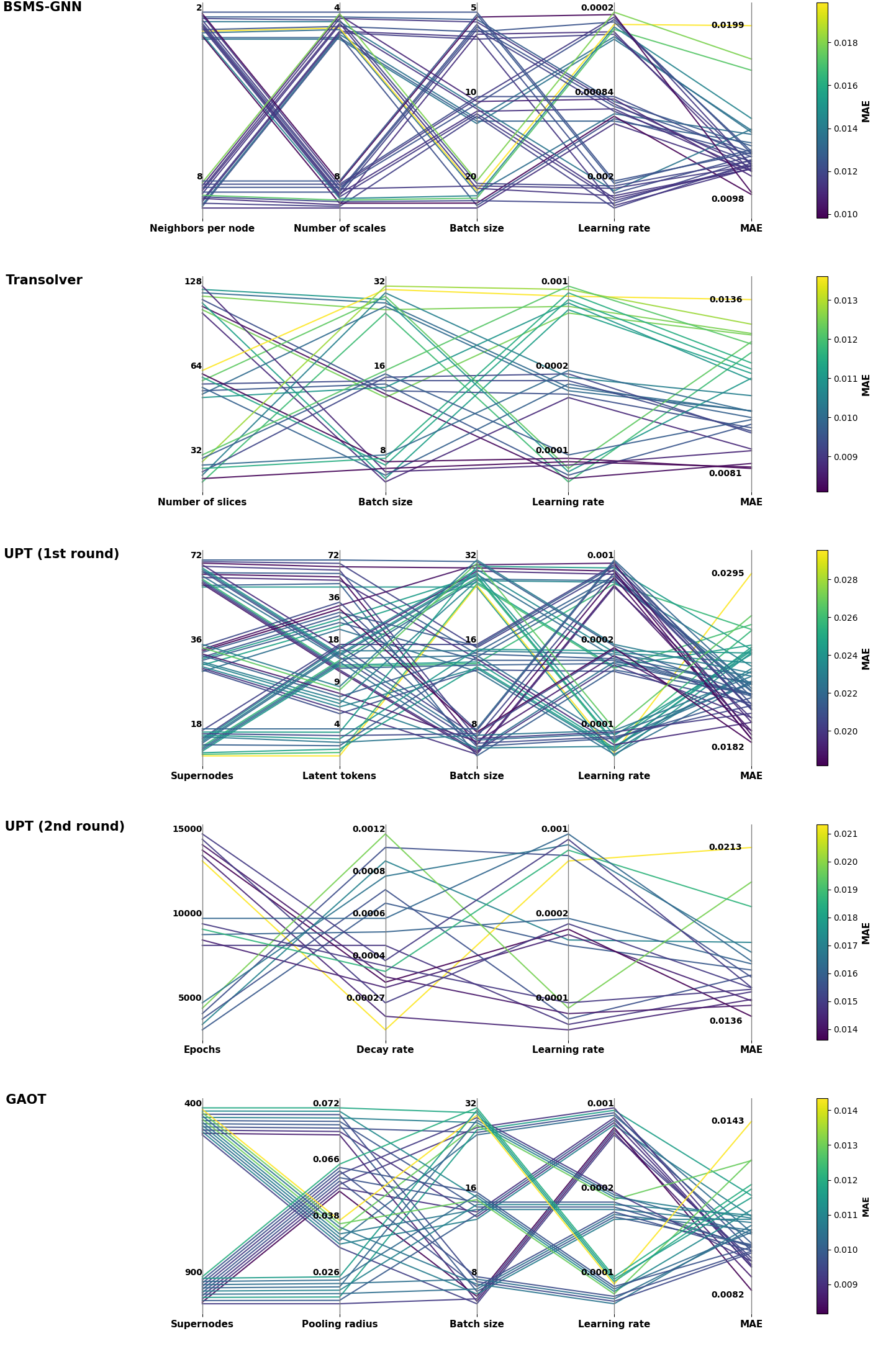}
\caption{Parallel coordinates plots visualizing the hyperparameter optimizations for the individual models. Each colored line corresponds to a specific combination of hyperparameters and the resulting MAE on the validation set.} \label{parallelcoords}
\end{figure}

The impact of the individual hyperparameters on each architecture is displayed in Figure \ref{parallelcoords}. For BSMS-GNN, the combination of the lowest max.\ learning rate and the highest batch size resulted in the four highest MAEs, giving these two hyperparameters the largest influence on the model's learning capabilities. On average, a batch size of $20$ amounted to an MAE of $0.013551$, whereas the lowest batch size of $5$ lead to an average MAE of only $0.011812$. Nonetheless, the lowest MAE was achieved for a batch size of $20$ and a max. learning rate of $0.00084$, which, against this backdrop, gives the impression of being a potential outlier. Regarding the numbers of neighbors and scales, instead, only a relatively moderate tendency of improved results for the higher number of scales $8$ (MAE $0.012013$) over $4$ scales (MAE $0.013028$) was observable on average, indicating that a deep multi-grid scheme and exchange between large amounts of nodes might not be necessary to capture the essential information in the dataset.

For Transolver, as well, the results were mostly influenced by the learning rate and the batch size: The combination of the lowest max.\ learning rate and the lowest batch size lead to three of the four lowest MAEs on the validation set, with the third lowest MAE resulting from the second lowest batch size. In comparison, the number of slices did not seem to crucially affect the performance with both the lowest and the highest MAE being achieved with the median number of $64$ slices and only a very marginal difference between the average MAEs attained for $32$ slices ($0.010652$) and $128$ slices ($0.010247$), hinting that all investigated values surpassed the number of physical states recognizable in the data samples.

The UPT architecture also reacted strongly to changes in the max.\ learning rate and the batch size, degenerating for the smallest considered learning rate and the largest considered batch size, respectively, during the first optimization round. Similarly, the model consistently performed worst for the lowest numbers of supernodes and latent tokens. For larger values of the latter parameters, the outcome was less conclusive. While the combination of both 72 supernodes and latent tokens lead to lower MAEs on average, the combination of 36 supernodes and latent tokens resulted in two out of the four best outcomes, including the overall lowest MAE. This hints once more that on the small-scale 2D airfoil dataset architecture sizes beyond a certain threshold might be redundant for extracting the required information. In the second optimization round, focused on the number of epochs, the learning rate and its decay rate, the results switched. With the linear learning rate decay during the first round replaced by exponential decay, UPT now performed better for the lowest max.\ learning rate of $0.0001$, attaining an MAE of $0.015568$ on average, than for the largest one of $0.001$ (MAE $0.017172$). Nonetheless, the lowest MAE was achieved for the same max.\ learning rate of $0.0002$ as during the first round. The second round also confirmed the positive effect of longer training times, as the MAEs attained by the instantiations trained for $15.000$ epochs - except for one outlier of $0.021335$ - consistently ranged among the lowest achieved values, benefiting from the additional weight updates.

Finally, the GAOT architecture revealed preferences for the smallest considered batch size and the largest considered max.\ learning rate with average MAEs of $0.009670$ and $0.009569$ attained for these two hyperparameters, respectively, compared to average MAEs of $0.011134$ for the largest batch size and $0.011162$ for the smallest max.\ learning rate. For GAOT also the impact of the combination of the number of supernodes and the pooling radius was clearly recognizable with an average MAE of $0.011406$ in the case of $400$ supernodes and a radius of $0.038$ in comparison to an MAE of $0.009582$ for $900$ supernodes and a radius of $0.066$. A potential explanation for this - in comparison to the corresponding hyperparameters in the remaining architectures - large influence could lie in GAOT's sampling strategy: The supernodes being selected away from the input points might increase the model's sensitivity to their number and pooling radius.

\subsection{Results} \label{results2d}

For each model we trained ten instantiations, employing the respective best hyperparameter setting according to the optimization procedure in Section \ref{hpopt}, for the prediction of the surface pressure on the full $498$ training samples of the 2D airfoil dataset. Subsequently, each instantiation was evaluated in terms of a comprehensive set of error metrics on the $100$ unseen test samples. The results in the form of the average errors of the ten instantiations per model as well as the corresponding $95\%$ confidence intervals are presented in Table \ref{Tab:tabRes2d}.

The results reveal a clear hierarchy of the considered models regarding all error metrics with increasing performance gaps between the architectures towards the lower end of the ranking. The highest accuracy is achieved by Transolver, followed tightly by BSMS-GNN, then GAOT and, somewhat behind, UPT. This hierarchy also applies with respect to the $R^2$-score, however, due to the nature of the $R^2$-coefficient the differences here appear marginal, making them difficult to interpret. We will thus exclude the latter metric from our following analysis.

The comparison between the two best-performing architectures comes out rather close: Transolver slightly surpasses BSMS-GNN on average with respect to each error metric taken into account. Its errors undercut the ones of BSMS-GNN by $10\%$ (MAE, RMSE) to $14\%$ (Rel. L2). The comparison of the associated confidence intervals fluctuates from metric to metric. While the intervals regarding the MAE, the RMSE and the relative $L^1$- and $L^2$-errors are separated by distances of $1.6$ (MAE) to $3.7$ (Rel. L2) times the size of the highest corresponding uncertainty, an overlap occurs between the two models' confidence intervals with respect to the MSE. % ($(0.001060, 0.001202)$ (BSMS-GNN) vs.\ $(0.000965, 0.001071)$ (Transolver)).

Similarly, GAOT does not fall too far behind BSMS-GNN, however, the gaps between these two models are slightly larger than between BSMS-GNN and Transolver. BSMS-GNN's average errors undercut the ones of GAOT with respect to every metric by $14\%$ (RMSE, Rel. L2) to $23\%$ (MSE). Furthermore, no overlaps are present between the confidence intervals for BSMS-GNN and GAOT, with the interval regarding the MSE %($(0.001060,\ 0.001202)$ (BSMS-GNN) vs.\ $(0.001281,\ 0.001503)$ (GAOT))
now lying apart by $0.7$ times the largest associated uncertainty. For the intervals with respect to the remaining error metrics, this distance even increases to between $3.0$ (RMSE) and $5.8$ (Rel. L1). We point out that, interestingly, while on the validation set after the hyperparameter optimization GAOT had achieved an MAE lower than BSMS-GNN and comparable to Transolver, it does not transfer this strong performance to the final evaluation on the test set. In fact, out of all the investigated architectures, GAOT is the only one which achieved a lower MAE during the hyperparameter optimization than on the test set after being trained on the full training data.

Finally, with average errors ranging $25\%$ (RMSE) to $52\%$ (Rel. L1) higher than the ones of GAOT, UPT clearly comes off the worst out of all investigated architectures. This also shows in terms of the confidence intervals, separated by $0.9$ (MSE), as well as between $4.9$ and $8.0$ (RMSE, Rel. L2, MAE, Rel. L1) times the largest corresponding uncertainty. Presumably as a consequence of having significantly higher errors than the remaining models, the architecture further stands out by the size of its confidence intervals with uncertainties $1.4$ (MSE) to $3.0$ (Rel. L1) times larger than GAOT, suggesting a high variance between the solutions predicted by its ten individual instantiations.

Overall, judging by the evaluation in Table \ref{Tab:tabRes2d}, Transolver, BSMS-GNN and GAOT emerge as a group of fairly well-performing architectures on the dataset. Within this group, while, in terms of a comparison of the pure attained error metrics, the performance advantages of Transolver over BSMS-GNN and BSMS-GNN over GAOT do not appear too severe, the associated confidence intervals do indicate a certain statistical relevance to the established hierarchy. Between the best-performing (Transolver) and the worst-performing (GAOT) model of the group, the $25\%$ (RMSE) to $37\%$ (MSE) higher errors scored by the latter architecture are also not to be neglected. UPT instead falls off in comparison to this group with error scores ranging significantly higher than each of the remaining models.

\begin{figure}
\centering
\includegraphics[scale=0.465]{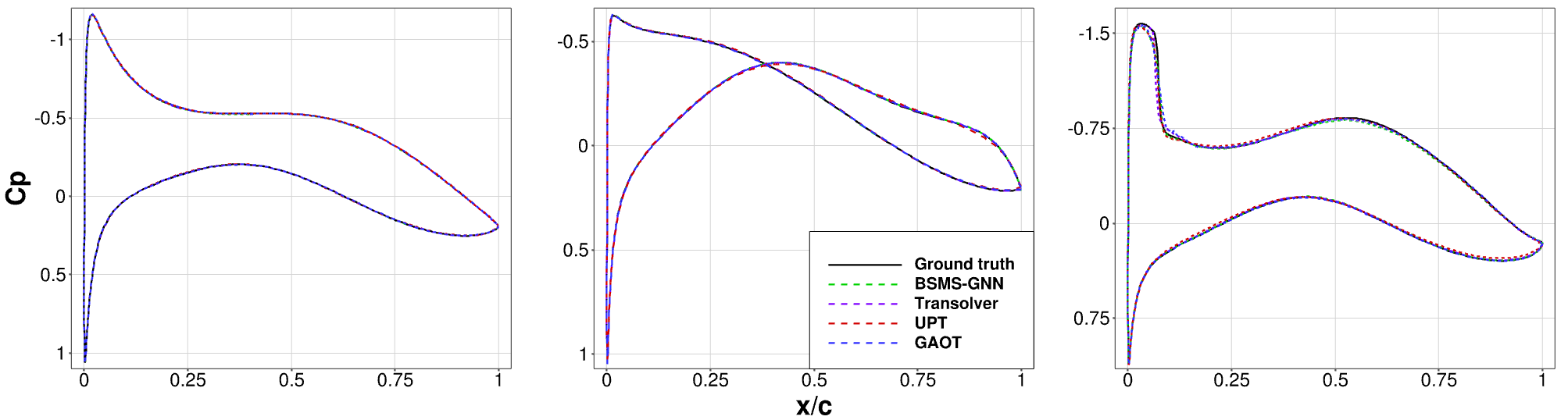} %[width=16cm,height=11cm]{solidinfluidnew}
%\hspace*{0.3cm}
\caption{Predictions of surface pressure distributions on three chosen test samples of the 2D airfoil dataset by UPT, GAOT, Transolver and BSMS-GNN. Each plot shows the ground truth (solid line) and the mean of the predictions (dashed lines) of ten fully trained instantiations of each model with optimal hyperparameters in accordance with Section \ref{hpopt}. The test samples correspond to the flow parameters $M:\ 0.490$, $AoA:\ 2.195$, $Re:\ 4.572*10^6$ (left), $M:\ 0.442$, $AoA:\ -2.484$, $Re:\ 2.901*10^6$ (middle) and $M:\ 0.674$, $AoA:\ 2.633$, $Re:\ 4.099*10^6$ (right).} \label{2dthreePredictions}
\end{figure}

Figure \ref{2dthreePredictions} shows the mean of the pressure predictions of the ten fully trained instantiations for each model in comparison to the ground truth on three chosen test samples. The first two columns represent two settings without the occurrences of any shocks, corresponding to Mach numbers of $0.490$ and $0.442$ as well as angles of attack of $2.195$ and $-2.484$, respectively. For the most part, these two tasks are solved sovereignly by all models with in particular little to no visible errors on the first sample. On the second sample, slight deviations become visible for UPT. The third column of Figure \ref{2dthreePredictions} corresponds to a setting involving a shock at a Mach number of $0.674$ and an angle of attack of $2.633$. Here, the predictions are of mixed quality. While the models remain mostly accurate away from the discontinuity, all of them reveal clear deviations from the target values in its vicinity. Remarkably, despite UPT displaying the by far greatest inaccuracy on the full test set, the errors close to the shock appear to circle around a similar level for all models with no advantage for any particular architecture. The significantly higher MAE of UPT ($0.020049$) in comparison to BSMS-GNN ($0.012211$), Transolver ($0.011719$) and GAOT ($0.012154$) on this sample instead appears to be caused by its slightly higher inaccuries distributed evenly over the remaining parts of the airfoil. Regarding the locating of the shock, the models exhibit different problems with all architectures except for Transolver placing its upper end too far to the left, whereas BSMS-GNN and UPT demonstrate a higher accuracy on its lower half.

\begin{figure}
\centering
\includegraphics[scale=0.69]{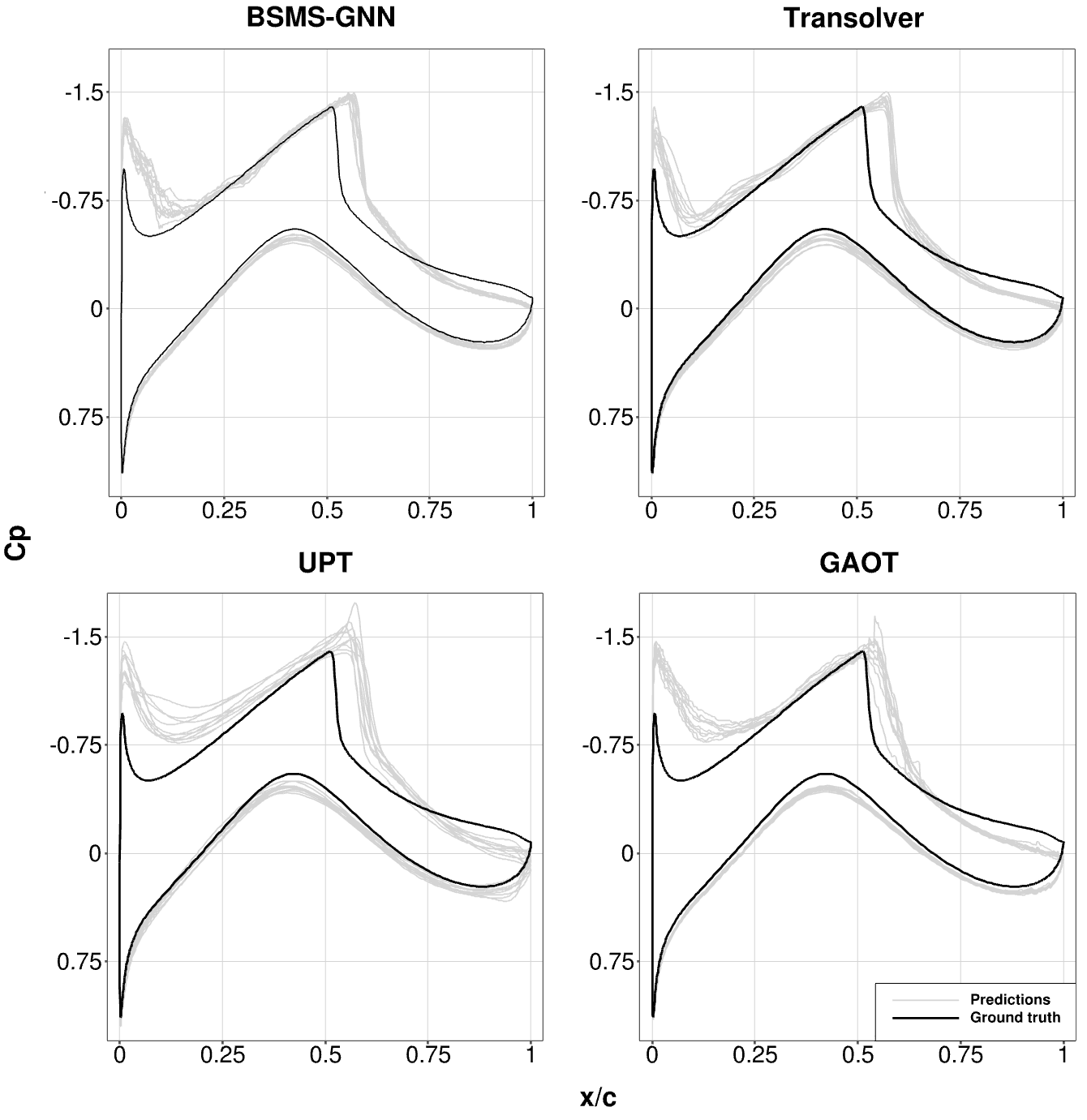} %[width=16cm,height=11cm]{solidinfluidnew}
%\hspace*{0.3cm}
\caption{Predictions and variance of surface pressure distributions by UPT, GAOT, Transolver and BSMS-GNN on the test sample corresponding to the flow parameters $M:\ 0.734$, $AoA:\ 3.758$, $Re:\ 4.701*10^6$. Each plot shows the ground truth (solid line) and the predictions (dashed lines) of ten fully trained instantiations of the respective model.} \label{2dVariance}
\end{figure}

Figure \ref{2dVariance} displays the performance of all ten individual instantiations of each model - and thereby the models' variance - on the test sample corresponding to a Mach number of $0.734$ and an angle of attack of $3.758$, on which all architectures attained their highest MAE ($0.126242$ (BSMS-GNN), $0.113807$ (Transolver), $0.168093$ (UPT), $0.162550$ (GAOT)). Here, the predictions of all models strongly deviate from the actual pressure values - not only close to the shock but along large parts of the entire geometry - with UPT's and GAOT's predictions lying the furthest from the ground truth. For the latter two architectures, the difficulties coming with this test case appear to cancel out their overall performance differences, with their accuracies roughly lying on par on this sample. Moreover, all models consistently displace the shock to the right of its actual location. We further point out that while - even if inaccurate - the predictions of Transolver and UPT remain fairly smooth, the predictions made by BSMS-GNN and GAOT show high-frequency oscillations around the shock. Interestingly, the smoothness of the predictions appears to be decoupled from the accuracy, as the smoothest predictions are created by the models with the highest (UPT) and the lowest (Transolver) MAE on the sample, respectively. A possible explanation of this behavior could lie in the architectural differences between the models: In contrast to Transolver and UPT, BSMS-GNN and GAOT heavily rely on GNN-components also in their later layers, where the local character of message passing might lead to fluctuations between the predictions in the individual domain points. Regarding the variance we observe that on this sample all models show rather strong deviations between the predictions made by their different instantiations. While for BSMS-GNN, Transolver and GAOT these deviations seem to move on a similar level, UPT, in accordance with its high uncertainties in Table \ref{Tab:tabRes2d}, exhibits the greatest variations, as is particularly visible to the left of the shock.

\begin{figure}
\centering
\includegraphics[scale=0.60]{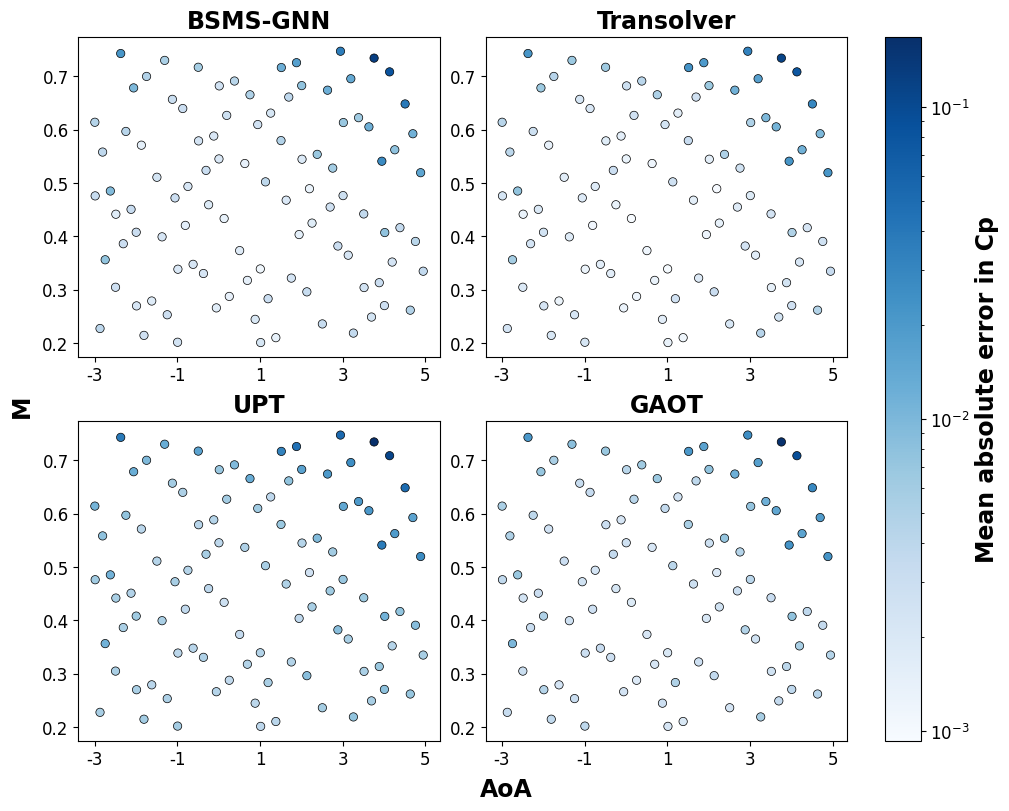} %[width=16cm,height=11cm]{solidinfluidnew}
%\hspace*{0.3cm}
\caption{Average of the MAEs attained by the ten fully trained instantiations of each model on all 100 test samples of the 2D airfoil dataset (log plot). The MAEs are displayed in dependence of the angle of attack $AoA$ and the Mach number $M$ of the samples, neglecting the influence of the airfoil geometry and the Reynolds number.} \label{2dMachAlpha}
\end{figure}

A local error analysis investigating the performance of the different architectures on the individual test samples is presented in Figure \ref{2dMachAlpha}, which plots the average MAE achieved by the ten instantiations of each model in dependence of two of the input features, namely, the angle of attack and the Mach number in the respective sample. Predictably, the models struggle the most in boundary cases with the highest errors being concentrated in the top right corner of the design space corresponding to especially high values of $AoA$ and $M$. We point out that, while very high Mach numbers appear to cause difficulties for most values of $AoA$, large angles of attack do not seem to significantly increase the MAE if $M$ remains small, presumably due to the absence of shocks in that regime. We further notice that, in accordance with our observations from the global error metrics in Table \ref{Tab:tabRes2d}, BSMS-GNN, Transolver and GAOT achieve visibly higher accuracies on the majority of samples than UPT, whereas the disparity within this group is significantly less distinct. In terms of the maximum MAE on the full test set, however, GAOT and UPT are roughly on par, since the performance differences appear to be nullified in a few of the more delicate test cases, as observed in Figure \ref{2dVariance}.

\begin{figure}
\centering
\includegraphics[scale=0.41]{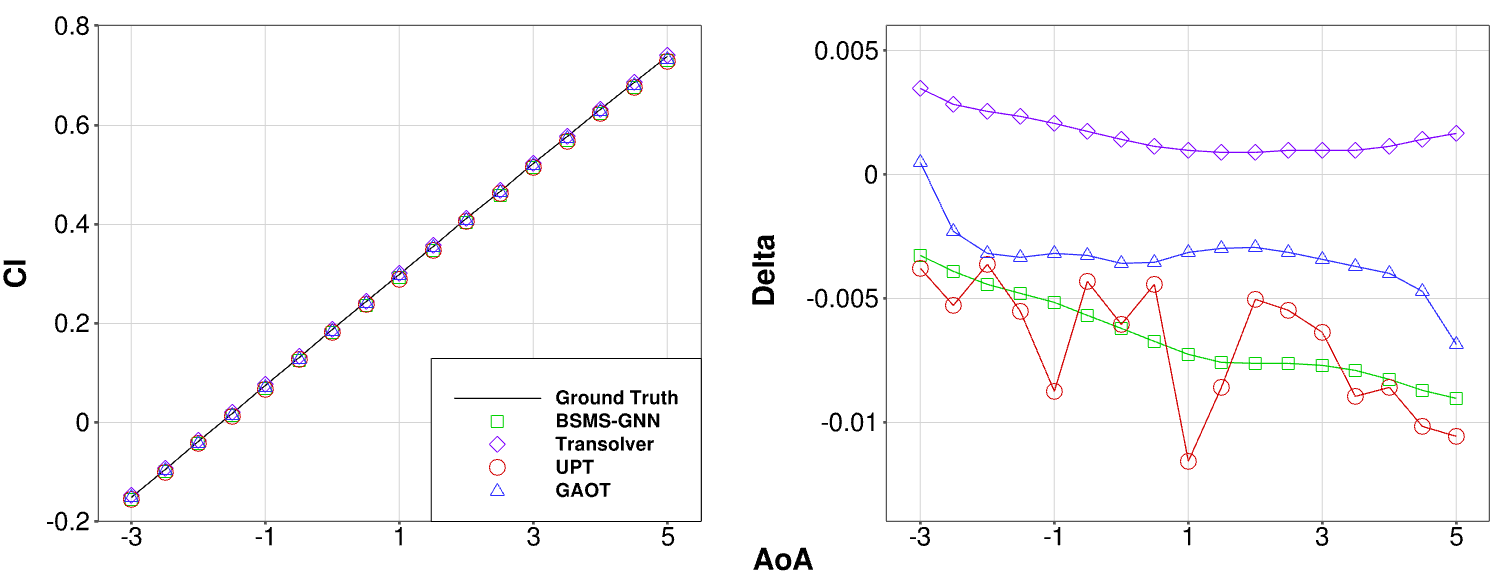} %[width=16cm,height=11cm]{solidinfluidnew}
%\hspace*{0.3cm}
\caption{Predictions of the lift coefficient $C_l$ of BSMS-GNN, Transolver, UPT and GAOT and difference $\Delta$ between the predicted values and the ground truth on $17$ additional datasamples corresponding to the original RAE2822 airfoil geometry at $M=0.2$, $Re=10^6$ and $AoA$ varying between $-3.0$ and $5.0$. The plotted values correspond to the mean of the $C_l$ predictions of the ten fully trained instantiations of each architecture, calculated from the respective pressure predictions.} \label{cloveralpha}
\end{figure}

Figure \ref{cloveralpha} visualizes the mean of the lift coefficients $C_l$ calculated from the pressure predictions of the ten instantiations of each model on $17$ additional datasamples corresponding to the original RAE2822 airfoil geometry, a Mach number of $0.2$, a Reynolds number of $10^6$ and variable angles of attack ranging from $-3.0$ to $5.0$. Expectedly, Transolver shows the least deviation from the ground truth with the highest precision attained for the medium values of $AoA$ lying in the center of the design space. Interestingly, Transolver's predictions consistently lie above the ground truth, whereas the remaining architectures show a clear tendency of underpredicting the lift coefficient. Moreover, UPT stands out by large jumps between $C_l$ values predicted for neighboring angles of attack, making its lift coefficient predictions the most inconsistent ones among the investigated models. From a practical point of view, however, the errors attained here are insignificant and all models demonstrate the capability of making satisfactory $C_l$ predictions.

\begin{figure}
\centering
\includegraphics[scale=0.6]{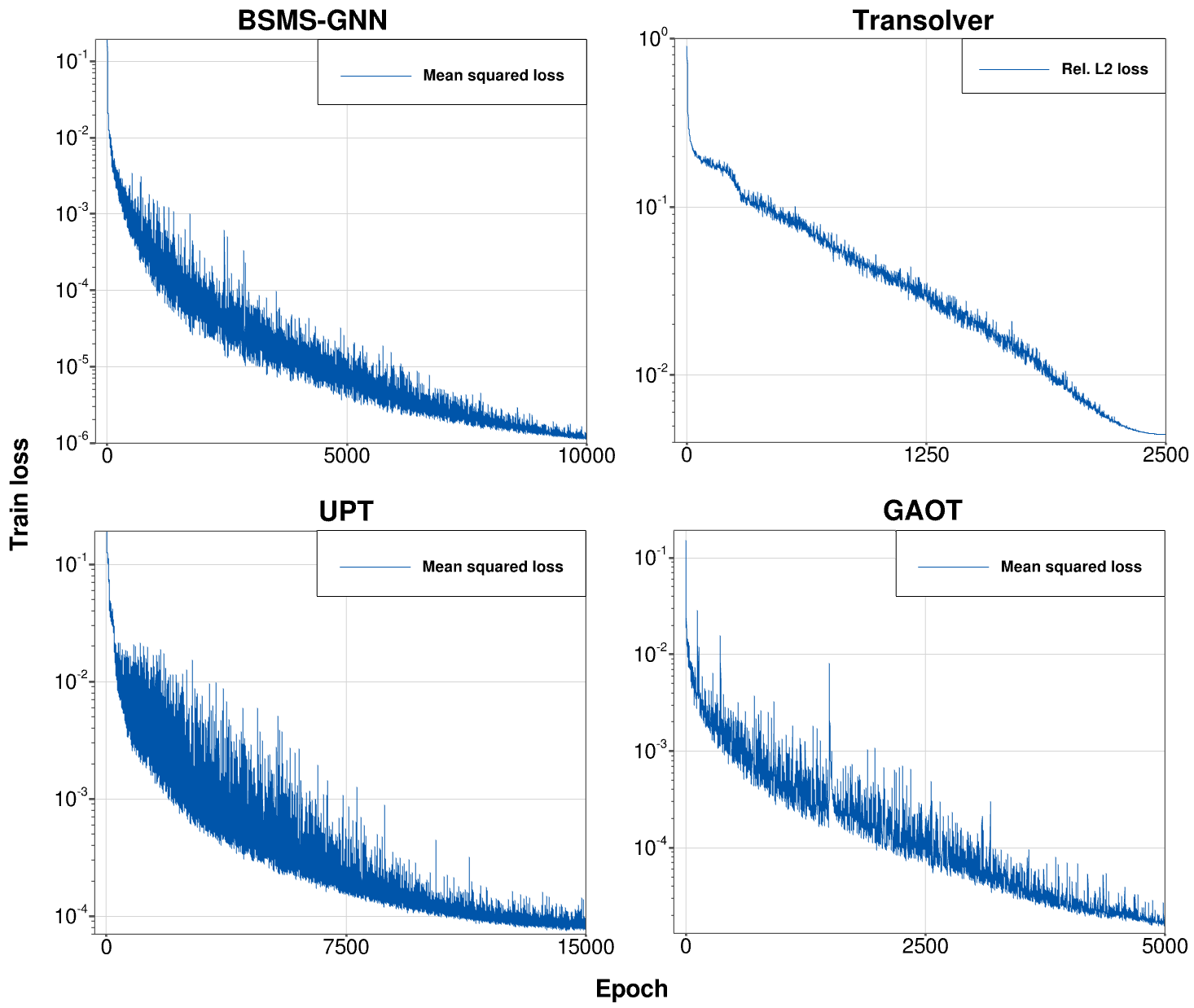}
\caption{Evolution of the training loss (log plot) of the individual models on the full 2D airfoil training set. Each curve corresponds to the loss of the respective model's instantiation achieving the one below median MAE in the ten repetition runs evaluated in Section \ref{results2d}. BSMS-GNN, UPT and GAOT were trained based on mean squared loss; for Transolver relative $L^2$-loss was used.} \label{lossCurves}
\end{figure}

Finally, a comparison of the computational efficiencies of the evaluated models on an NVIDIA Quadro P2200 GPU with 5GB memory is given in Table \ref{Tab:tabRes2deff}. What stands out is that, in the form of Transolver and BSMS-GNN, the two models which achieved the highest predictive accuracy are also the most lightweight ones, with BSMS-GNN having almost 4 times fewer parameters than UPT and 8 times fewer parameters than GAOT. With regards to the training time, however, the situation is more convoluted. We first notice that each model was trained for a different amount of epochs. For UPT a total of $15,000$ epochs was used in accordance with the results of the hyperparameter optimization in Section \ref{hpopt}, whereas BSMS-GNN was trained for $10,000$ epochs in accordance with previous experiences with the model. Transolver and GAOT, instead, were trained for $2,500$ and $5,000$ epochs, respectively, as these numbers had shown to be sufficient to achieve a fully trained state in initial experimental runs. Loss curves displaying the training progress of individual instances of all four models are pictured in Figure \ref{lossCurves}. The longest time per training epoch ($5.33$s) is taken up by GAOT. Besides the high parameter count, this appears to be related to GAOT being equipped with a total of $900$ supernodes. In comparison, UPT, which relies on only $36$ supernodes, completes its training epochs in only $2.43$s on average. Nonetheless, GAOT's overall training time of $7$h $24$m $1$s lies below the ones of both UPT ($10$h $8$m $30$s) and BSMS-GNN ($9$h $44$m $15$s), due to the lower amount of epochs the model takes to be fully trained. The lowest overall training time is achieved by Transolver, which requires $2$h $42$m $43$s of training to attain the highest accuracy in the experiment. Furthermore, neglecting the training procedure, we observe that GAOT in fact achieves an inference latency comparable to Transolver and UPT ($\leq 10$ms), while only BSMS-GNN requires slightly more time for individual predictions ($\leq 15$ms). However, in comparison to a full CFD run, this is negligible.

\section{Surface pressure prediction on 3D NASA CRM aircraft configuration} \label{nasacrm}

Our second experiment extends our previous investigations to the 3D setting. More specifically, we are interested in the models' capabilities to predict the surface pressure distribution on the NASA CRM civil aircraft configuration. To this end, we consult the dataset introduced in \cite[Section II.D]{rae2822}, encompassing the flow data on this configuration under 149 different operational conditions. With $454,404$ surface points per sample, this dataset allows for an investigation of the models' aptitude for handling complex industrial-scale data. At the same time, the size of only 149 samples is ideally suited to simulate data scarcity, which represents another classical obstacle in industrial problems. In contrast to the 2D airfoil dataset, the general geometry is fixed across all samples in the NASA CRM dataset and varies only in terms of the inboard aileron deflection angle, the outboard aileron deflection angle, the elevator deflection angle and the horizontal tailplane (HTP) angle, which are treated as global parameters instead of changing coordinates. In each sample, these configuration parameters are further accompanied by two global flow parameters, namely, the Mach number $M$ and the angle of attack $AoA$. These six parameters, which were chosen according to a Halton sequence, form the operational conditions of the 149 samples in the dataset. In dependence of them, the pressure coefficient $C_p$ and the skin friction coefficients $C_{f_x}$, $C_{f_y}$, $C_{f_z}$ were originally calculated, via the CFD solver DLR TAU, as solutions to the RANS equations in conjunction with the Spalart-Allmaras turbulence model. In our experiment, as in our corresponding study on the 2D airfoil dataset, we do not take into account the skin friction coefficients. Instead, we train the models to predict the surface pressure in dependence of the spatial coordinates, the surface normals and the operational conditions. As in the 2D case, we normalize all input - but not output - features via min-max scaling. All input and target quantities are listed in Table \ref{Tab:tabCRM}; the stated lower and upper bounds correspond to their values before the scaling.

Out of the overall 149 samples, 75 samples are intended for training, 30 are intended for validation and 44 are intended for testing. However, due to the excessive training times on this large-scale dataset (cf.\ Table \ref{Tab:tabRes3deff}), we forewent a hyperparameter optimization in this case and instead directly trained the models on the 105 samples from the combined training and validation sets, before evaluating them on the remaining 44 test samples. The hyperparameter settings we employed in this process were chosen on the basis of our experiments in the 2D case, for the precise settings we refer to Section \ref{setup} in the appendix. For the same reason, we abstained from doing repetition runs and instead assessed a single instantiation per model. The high memory usage caused by the sheer size of the individual samples further prevented us from training the original Transolver model on this dataset. Therefore, in this experiment we instead used Transolver++, a highly parallelized extension of Transolver \cite{transolver++}, which was specifically designed to tackle the hardships coming with industrial-scale data.

\subsection{Results} \label{resultsNASACRM}

We trained each of the four models of interest on the full $105$ training and validation samples of the NASA CRM dataset, before evaluating them on the remaining $44$ test samples. The resulting error scores are summarized in Table \ref{Tab:tabRes3d}.

As in the 2D case, the results establish an unambiguous ranking of the models' performance on the NASA CRM dataset regarding all error metrics with two main differences: First, the ranking of the two best-performing architectures has been reversed with now BSMS-GNN scoring the lowest errors throughout all metrics. Second, the performance gaps between the individual architectures are significantly higher, splitting the selection into a group of two decently performing models (BSMS-GNN, Transolver++) and a group of two models exhibiting pronounced to severe difficulties (GAOT, UPT) on this dataset. While both BSMS-GNN and Transolver++ demonstrate a respectable accuracy, the increased performance differences are already noticable within this group. With respect to RMSE, Rel. L2, Rel. L1 and MAE, BSMS-GNN outperforms Transolver++ by $32\%$ to $34\%$ lower errors, while coming out ahead by a whole $59\%$ in terms of the MSE. The architectures of the second group, lead by GAOT, fall off even more drastically in comparison. With an MAE of $0.023976$, GAOT, despite apparent struggles, still shows the capability of approximately learning the correct solutions. However, errors ranging between $1.8$ (RMSE) and $2.4$ (MSE) times higher than the ones of Transolver++ do not place GAOT anywhere close to the first group. Finally, UPT yet again exceeds the errors of GAOT by factors between $1.8$ (RMSE, Rel. L2) and $3.0$ (MSE). The attained MAE barely attests the architecture any capacity of learning the required information from the data and UPT performs noticably worse even in terms of the $R^2$-score. We point out that while, due to the lack of repetition runs, the results presented in Table \ref{Tab:tabRes3d} do not allow us to make any statistical statement, their relevance is heightened by the fact that the errors scored by the individual models are separated by gaps multiple times larger than the size of the respective confidence intervals in the case of the 2D airfoil dataset, cf.\ Table \ref{Tab:tabRes2d}.

\begin{figure}
\centering
\includegraphics[scale=0.55]{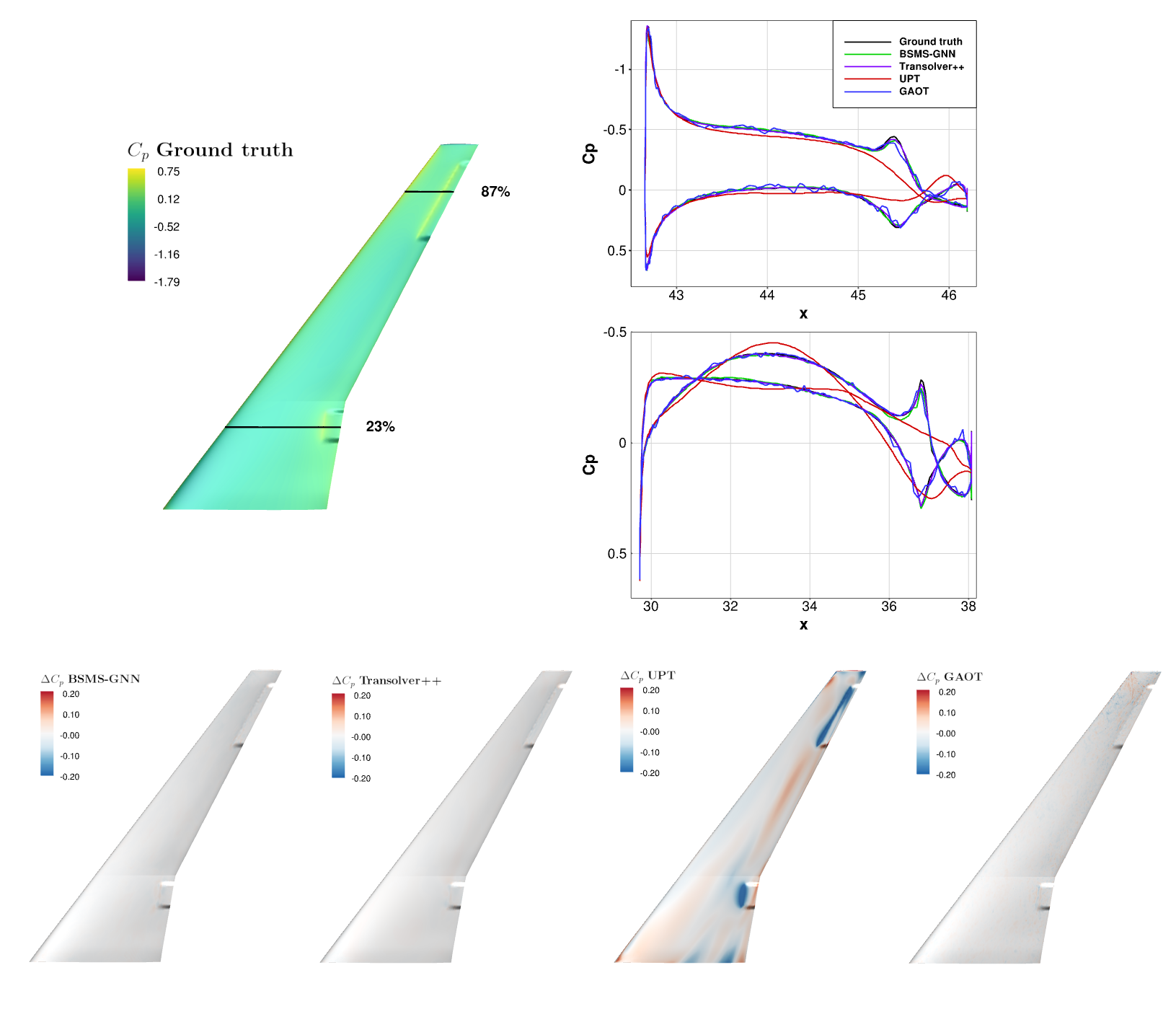}
\caption{Predictions and errors of all models on a test sample corresponding to a Mach number of $0.547$, an angle of attack of $1.265$, an outboard aileron angle of $14.240$, an inboard aileron angle of $11.603$, an elevator angle of $9.835$ and an HTP angle of $-0.864$. The upper left plot shows the ground truth pressure distribution on the right wing of the aircraft configuration, the two plots to its right display the models' predictions in two cross sections on the level of the outboard and inboard aileron, respectively. The bottom row shows the error $\Delta C_p$ (prediction minus ground truth) between all models' predictions and the target values.} \label{predictionsNASACRMSample119}
\end{figure}

\begin{figure}
\centering
\includegraphics[scale=0.55]{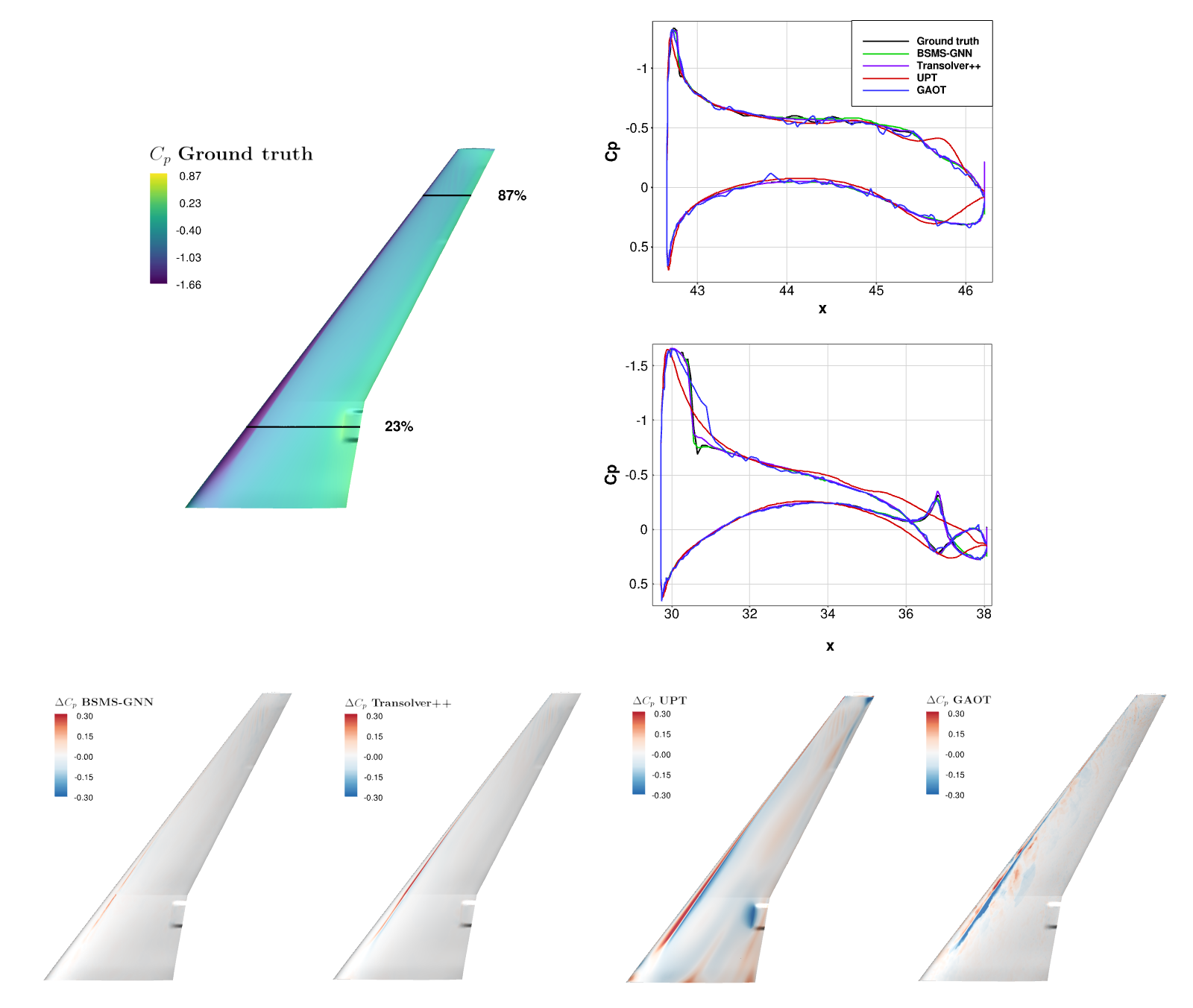}
\caption{Predictions and errors of all models on a test sample corresponding to a Mach number of $0.722$, an angle of attack of $3.457$, an outboard aileron angle of $1.360$, an inboard aileron angle of $-12.420$, an elevator angle of $2.893$ and an HTP angle of $0.959$.} \label{predictionsNASACRMSample112}
\end{figure}

\begin{figure}
\centering
\includegraphics[scale=0.55]{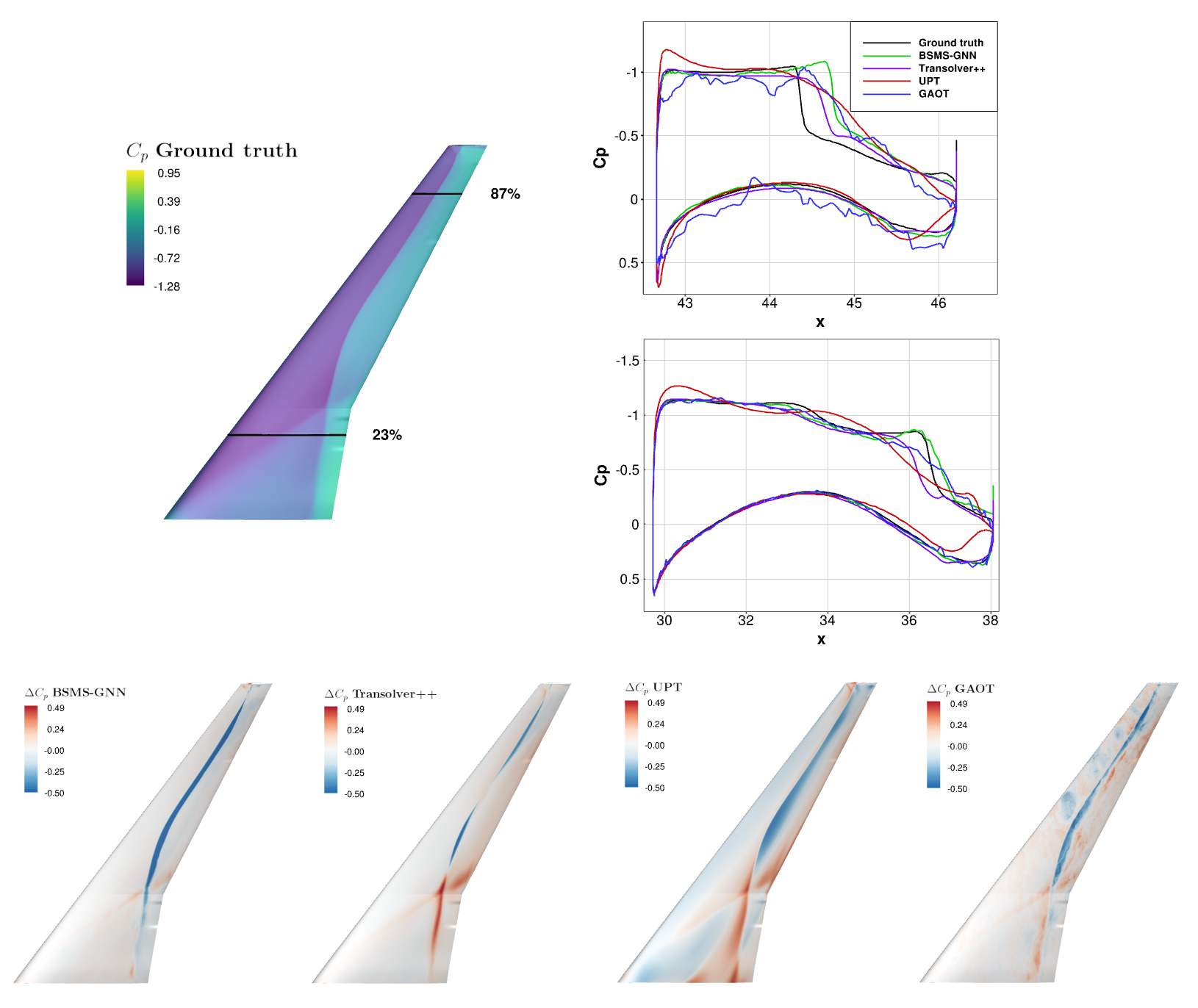}
\caption{Predictions and errors of all models on a test sample corresponding to a Mach number of $0.877$, an angle of attack of $4.383$, an outboard aileron angle of $1.648$, an inboard aileron angle of $2.274$, an elevator angle of $-9.654$ and an HTP angle of $-1.740$.} \label{predictionsNASACRMSample142}
\end{figure}

On three chosen test samples, Figures \eqref{predictionsNASACRMSample119}--\eqref{predictionsNASACRMSample142} provide a visualization of the models' predictive capabilities, displaying the respective ground truth pressure distribution, each model's deviation $\Delta C_p$ (prediction minus ground truth) from the target values as well as each model's prediction along two cross sections at the level of the outboard and the inboard aileron on the right wing of the aircraft configuration. We remark that we have restricted these plots to the right wing as this appears to be where the models' highest errors are concentrated.

With a Mach number of $0.547$ and an angle of attack of $-1.265$, the sample depicted in Figure \eqref{predictionsNASACRMSample119} corresponds to a rather simple setting without the occurrence of any shocks but relatively large inboard and outboard aileron angles of $-11.603$ and $14.240$, respectively. Both BSMS-GNN and Transolver++ handle this sample with barely any errors visible in the pressure distribution across the full wing. The cross section plots provide a similar picture with both models' predictions aligning closely with the ground truth except for noticable deviations around the disruptions caused by the ailerons. The predictions made by GAOT appear to follow the general course of the ground truth pressure distribution with similar accuracy, however, they are diluted by high-frequency oscillations as we had already observed them in the 2D case, cf.\ Figure \ref{2dVariance}. In particular on the outboard section of the wing, this leads to visibly higher errors than for BSMS-GNN and Transolver++. UPT instead showcases evident errors across large parts of the whole wing. Already away from the ailerons, the architecture exhibits higher inaccuracies than the remaining models, even though it still manages to capture the overall shape of the pressure distribution in both cross section plots. The impact of the ailerons, however, appears to be entirely ignored by UPT, leading to a severe underprediction of the pressure in these regions.

The sample in Figure \eqref{predictionsNASACRMSample112} corresponds to a Mach number of $0.722$ and an angle of attack of $3.457$, resulting in a clearly observable shock present along the leading edge. While the inboard aileron angle with a value of $12.420$ in this sample is again rather large, the outboard aileron angle of $1.360$ shows a negligible effect on the pressure distribution. The shock is predicted with different precision by the individual models with the most favorable results stemming once more from BSMS-GNN and Transolver++. Both architectures catch the position of the shock with high accuracy and only showcase a relatively minor overprediction of the target pressure to its right side on the inboard section. Away from the discontinuity, they show strong compliance with the ground truth including only slight deviations at the inboard aileron. GAOT, in comparison, struggles more with the shock. While on the level of the outboard aileron, where the discontinuity is less distinct, the model remains fairly accurate, its predictions on the level of the inboard aileron lie visibly below the ground truth to the left and severely above the target values to the right of the discontinuity. On the remaining parts of the wing geometry, GAOT again presents itself competitive with BSMS-GNN and Transolver++ in terms of reproducing the general shape of the pressure distribution, but reduces its accuracy by its oscillations, which are particularly pronounced on the outboard section. UPT stands out by the least steep prediction in the inboard area, barely capturing the discontinuity at all, as well as the predictions lying the furthest below the target values to the left of the shock out of all models. Moreover, UPT, again, does not show any reaction to the impact of the inboard aileron. We point out that on this sample, the target pressure distribution further exhibits some slight waves close to the tip of the wing. These waves, however, give away the impression of being unphysical artifacts resulting from the CFD calculation and are ignored by all models.

Finally, in Figure \eqref{predictionsNASACRMSample142}, we see the predictions on the sample which, with a combination of a high Mach number of $0.877$ and a high angle of attack of $4.383$, results in the highest MAEs ($0.032654$ (BSMS-GNN), $0.045149$ (Transolver++), $0.112510$ (UPT), $0.060934$ (GAOT)) for all models on the entire test set. Both the inboard aileron angle of $2.274$ and the outboard aileron angle of $1.648$ instead are rather moderate. The sample features a $\lambda$-shock, composed of one prominent shock, ranging vertically from the inboard end to the outboard end of the wing as well as one less pronounced discontinuity moving diagonally from the leading to the trailing edge. While BSMS-GNN, Transolver++ and GAOT manage to achieve a somewhat passable prediction of the diagonal shock, all models display severe difficulties in dealing with the vertical one. This holds true especially in the outboard region, where this shock is consistently placed too far downstream, leading to a strong overprediction of the pressure by all architectures. Remarkably, in the outer region of the wing, BSMS-GNN appears to handle the positioning of the shock the worst, however, due to higher accuracy in the inner region and away from the shock, still undercuts the other models in terms of the MAE. On this sample, all models further display visible deviations from the ground truth on large parts of the wing away from the shock, with the greatest inaccuracies being seen in the predictions of UPT. In the outboard area, however, UPT's inaccuracies appear to be matched in terms of magnitude by GAOT's large oscillation-induced errors. Furthermore, on this sample, also BSMS-GNN's oscillations, which we had already witnessed in the 2D case (cf.\ Figure \eqref{2dVariance}), come to light again. These oscillations, however, differ from the ones of GAOT in their significantly lower amplitude.

\begin{figure}
\centering
\includegraphics[scale=0.60]{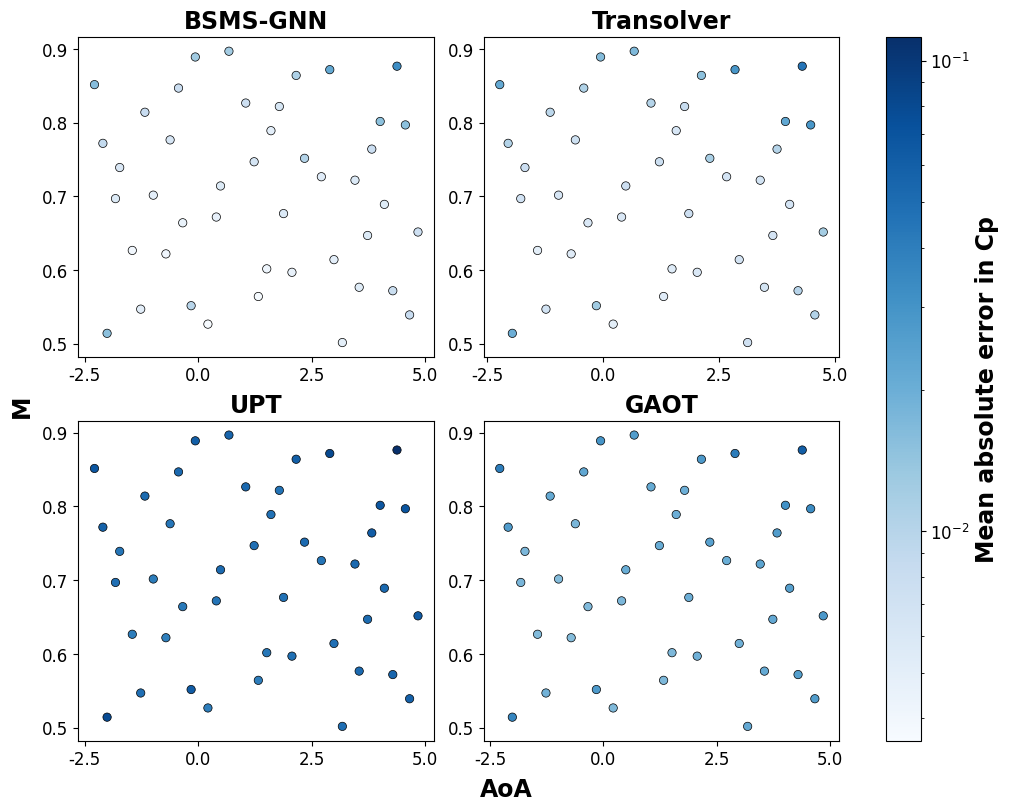} %[width=16cm,height=11cm]{solidinfluidnew}
%\hspace*{0.3cm}
\caption{Mean absolute errors of the fully trained instantiations of BSMS-GNN, Transolver++, UPT and GAOT on all $44$ test samples of the NASA CRM dataset (log plot); plotted in dependence of the angle of attack $AoA$ and the Mach number $M$.} \label{machalphanasacrm}
\end{figure}

Figure \ref{machalphanasacrm} displays the mean absolute errors attained by BSMS-GNN, Transolver++, UPT and GAOT on the individual test samples of the NASA CRM dataset in dependence of the respective Mach numbers and angles of attack. We point out that, despite the larger average MAEs attained by all architectures on the full test set, the maximum MAEs lie significantly below the highest errors of the corresponding models in the 2D case. Apart from that, for BSMS-GNN and Transolver++, a similar picture as in the 2D case (cf.\ Figure \ref{2dMachAlpha}) emerges: Both models reveal their greatest difficulties in boundary cases, with the highest errors resulting from large angles of attack and, with even greater weight, large Mach numbers. In accordance with the observations from Table \ref{Tab:tabRes3d}, however, the errors of Transolver++ are visibly higher than for BSMS-GNN. The rather high MAE scored by GAOT on the full test set also transfers to the local error analysis with relatively large errors throughout all samples. While the architecture struggles the most in the same boundary cases as BSMS-GNN and Transolver++, it appears to distribute its errors slightly more evenly across the test set. Indeed, the ten largest MAEs attained by Transolver++ are, on average, $3.0$ times larger than the average MAEs scored on the remaining test samples, whereas for GAOT this factor decreases to $1.7$. For UPT, this effect yet intensifies: While UPT clearly exhibits the largest errors across the entire test set, its average MAEs on its ten most problematic samples is only $1.5$ times larger than on the remaining ones.

\begin{figure}
\centering
\includegraphics[scale=0.6]{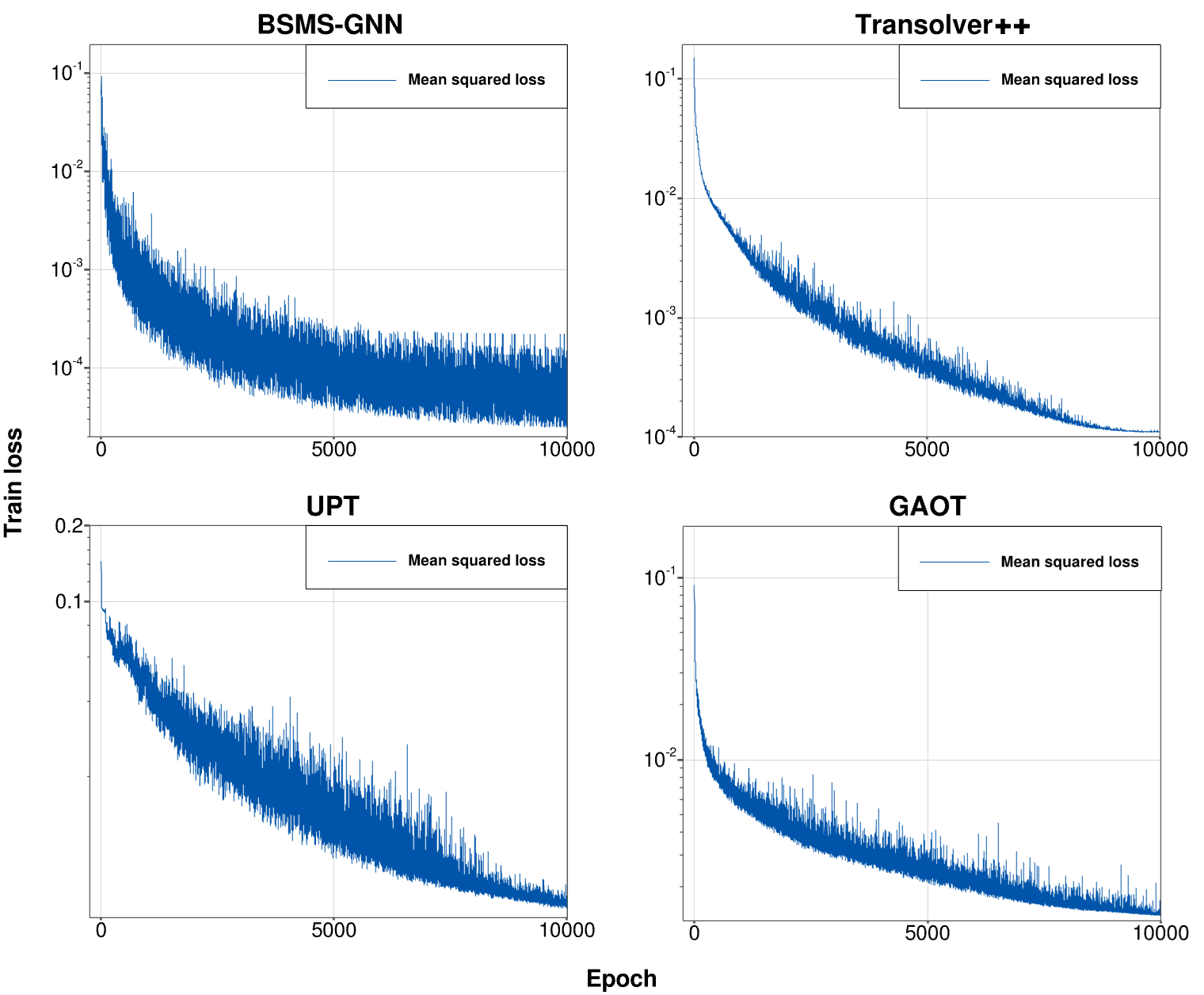}
\caption{Evolution of the training loss (log plot) of the individual models on the NASA CRM dataset. For all models mean squared loss was used as the loss function.} \label{lossCurvesNASACRM}
\end{figure}

Lastly, Table \ref{Tab:tabRes3deff} provides an overview of the computational efficiency of the trained instances of the investigated models on an NVIDIA H100 GPU with 80GB HBM3. We point out that, in this experiment, in order to ensure a fair comparison, we trained all models for a total of $10,000$ epochs, guaranteeing also BSMS-GNN and Transolver++ sufficient time to learn the required information from the intricate data. We also remark that we here trained BSMS-GNN, UPT and GAOT with a batch size of $1$ in an attempt to further improve the models' learning capabilities by increasing the frequency of the parameter updates. Transolver++, instead, was trained with a batch size of $4$, in order to keep its training time feasible. While the parameter counts of BSMS-GNN, UPT and GAOT increased slightly relative to their corresponding instantiations in the 2D case, Transolver++ stands out by more than twice as many learnable parameters in comparison to the instantiations of Transolver in the 2D setting, cf.\ Table \ref{Tab:tabRes2deff}. This large number can be attributed mainly to the increased hidden dimension in the attention mechanism in the default settings of Transolver++, cf.\ Table \ref{Tab:tabHPsTransolver} in the appendix. The increase of the weight is also reflected in the duration of the training, with Transolver++ taking the longest time ($58.38$s) for a single epoch on average, resulting in an overall training time of $6$d $18$h $10$m. As a close second, UPT comes in with $55.88$s per training epoch and a consequential wall-clock time of $6$d $11$h $13$m $20$s. While on the 2D airfoil dataset UPT had proved to be significantly faster than the remaining architectures, its slow training here might be explained by the severe increase of supernodes from $36$ to $20,480$, cf.\ Table \ref{Tab:tabHPsUPT}. In comparison, with $37.30$s and $37.73$s, respectively, BSMS-GNN and GAOT perform the best in terms of the time per training epoch. This is particularly remarkable in the case of BSMS-GNN, which therefore achieves the highest accuracy on the NASA CRM dataset within the lowest wall-clock time ($4$d $7$h $36$m $40$s). In the case of GAOT, which had required the longest time per epoch on the 2D airfoil dataset, it is interesting to notice that the architecture finishes its training epochs considerably faster than UPT, despite still having a higher parameter count and using more supernodes ($131,072$), cf.\ Table \ref{Tab:tabHPsGAOT}. A potential explanation lies in the fact that, while for UPT we employed $569$ times as many supernodes than in the 2D case, this factor only amounts to $146$ for GAOT. Loss curves showcasing the evolution of the training loss for all four models are displayed in Figure \ref{lossCurvesNASACRM}. Finally, with regards to the inference latency, we note that, relative to each other, BSMS-GNN, Transolver++ and GAOT behave roughly in accordance with their times per training epoch, while only UPT requires a significantly higher time for making individual predictions.

\section{Subjective assessment of the models' handling}

In addition to our empirical analysis, we provide a brief subjective judgement of the difficulties in the handling of the investigated models based on our experiences made in Sections \ref{rae2822exp} and \ref{nasacrm}. We stress that this assessment reflects our personal experiences rather than rigorous scientific findings. Nonetheless, we deem these insights valuable for a comprehensive evaluation of the architectures' suitabilities for use as effective surrogate models.

The different models display individual levels of complexity, with GAOT standing out as the most intricate. While its sophisticated architecture offers a variety of hyperparameters to tune in order to get the best out of the model's performance, this comes with the downside of increased difficulty in handling. The wide range of possible settings complicates the identification of the most efficient way to improve performance in the face of poor results. For instance, in order to keep within a reasonable timeframe, we felt compelled to restrict our hyperparameter optimization to GAOT's encoder. Additionally, the extensiveness of the model's code makes it difficult to get a clear picture of its functionality. In particular, modifications to the code, which can e.g.\ be necessary to process data in a format other than that intended by the authors can easily lead to errors and, as a result, a drastic deterioration in performance. Viewed from a different angle, however, the multitude of options in GAOT's set up could mean that the model's actual capabilities exceed what we have been able to observe in Sections \ref{results2d} and \ref{resultsNASACRM}, due to the range of possibilities which we had not the time to explore.

At the other end of the spectrum we have Transolver(++), which stands out for its simplicity: With its easy to understand architecture and its rather minimalistic code, the model can be quickly set up in a straight-forward way for custom problems. In our experiments, this became evident from the fact that very positive results were achieved even in initial experimental runs relying largely on the model's default settings. The model generally displayed a strong robustness, delivering consistently favorable outcomes throughout the runs in both the 2D and the 3D case. Transolver's plain transformer based design further bears the advantage that, in the event of hyperparameter optimization, few questions arise regarding the choice of hyperparameters and their values. In particular, the lack of GNN components is noticeable in this respect, greatly facilitating the hyperparameter selection as the construction of graphs is not required.

Finally, both BSMS-GNN and UPT fall somewhere in the middle between Transolver and GAOT in terms of handling. While their detailed architectures do not match the simplicity of Transolver, their designs are clear and the number of choices to be made in the selection of their hyperparameters remains manageable, allowing for a relatively straight-forward set up of both models. Given that the BSMS-GNN model has been previously used extensively within our working group it is only fair to note that the experience with this specific architecture exceeds the experience present with all other three architectures.

\section{Conclusion}

In this article, we assessed four state-of-the-art operator learning architectures - BSMS-GNN, Transolver(++), UPT and GAOT - on their applicability in aircraft aerodynamics. We evaluated these models in terms of their aptitude for learning surface pressure distributions in two different scenarios: First, on a 2D airfoil dataset, we examined their ability to predict the pressure distributions around diverse two-dimensional airfoil geometries for varying global flow parameters. Second, on the NASA CRM dataset, we tested their predictive capabilities on an industrial-scale three-dimensional aircraft configuration under changing operational conditions.

On the 2D airfoil dataset we began our investigation with a hyperparameter optimization, attempting to drive each architecture to its peak performance. Our observations indicated that all models mainly react to changes of the learning rate and the batch size used during the training procedure. The impact of hyperparameters related directly to the individual models, instead, appeared to be moderate, hinting that complex large-scale architectures might be not necessary for extracting the required information from the 2D airfoil data. In the subsequent evaluation of the optimized models, a clear hierarchy of the architectures became apparent. While all architectures demonstrated adequate learning capabilities on the dataset, the highest accuracy was achieved by Transolver, followed by BSMS-GNN, GAOT and then UPT. The results were statistically affirmed by rather large gaps between the corresponding confidence intervals. A local errors analysis revealed that for all models the greatest difficulties arise from similar test cases close to the boundaries of the design space, typically corresponding to large Mach numbers and large angles of attack. The performance differences between the models appeared to not be caused by degeneration around discontinuities, but rather by slight discrepancies distributed across the entire airfoil geometries. Also regarding the predictions of the lift coefficient, all models demonstrated a satisfactory precision. On the industrial-scale NASA CRM dataset, the performance differences between the architectures increased. Again, adequate results were attained by BSMS-GNN and Transolver++ with BSMS-GNN achieving the highest accuracy this time. GAOT, while still demonstrating the tendency to approximate the correct solutions, fell significantly behind with respect to all error metrics and UPT barely proved any learning capability on the intricate dataset. These large differences already became visible on rather simple test samples in the regime of low Mach numbers and angles of attack and became even more pronounced in the presence of shocks.

In terms of computational efficiency, it is noteworthy that in both the 2D and the 3D setting there was a clear trend of the best results being attained by the most lightweight models, with in particular the highest accuracies being achieved by the models requiring the least training time, respectively. These insights were further supplemented by the observations that the same models, particularly Transolver, also stand out for their rather straightforward handling.

Overall, we observed promising results for some models in both experiments. In the 2D setting, all models showcased the ability to make decent predictions, whereas in the 3D case BSMS-GNN and Transolver(++) pulled ahead, demonstrating superior learning capabilities. With further consideration of their computational efficiency and simple handling, our findings in particular highlight the latter two architectures as promising surrogate models. As more and more architectures keep getting published, future work may include extensions of our studies to new models; especially successors to the already investigated models such as Transolver-3 \cite{transolver3} or AB-UPT \cite{abupt} appear to be worth exploring. Also different approaches to operator learning, such as CNN-based neural operators like DoMINO \cite{domino} or foundational models like POSEIDON \cite{poseidon}, may be of interest for future work.

\begin{comment}
\section*{Acknowledgement}
The authors gratefully acknowledge the scientific support and HPC resources provided by the German Aerospace Center (DLR). The HPC system CARA is partially funded by ‘Saxon State Ministry for Economic Affairs, Labor and Transport’ and ‘Federal Ministry for Economic Affairs and Climate Action’.
\end{comment}

\appendix

\section{Appendix} \label{A}

\subsection{Experimental set ups} \label{setup}

We provide an overview of the hyperparameter choices we used in the models' set up and the training procedures during our experiments in Sections \ref{rae2822exp} and \ref{nasacrm}. Unless stated otherwise, these choices correspond to the default and exemplary settings in the publicly available code and the models' authors' suggestions in the accompanying articles, except for the hyperparameters which were optimized in our investigations on the 2D airfoil dataset in Section \ref{rae2822exp}. For each of the latter hyperparameters, we here state the respective optimum value as used in the final evaluations of the optimized models in Section \ref{results2d}. For the corresponding hyperparameters in the setting of the NASA CRM dataset, we either adopted the values from the 2D setting or adjusted them, to the best of our knowledge, based on our experiences during the optimization procedure. In the latter case, we explicitly comment on our choice in the following. 

Our hyperparameter choices for the set up and the training process of BSMS-GNN are listed in Table \ref{Tab:tabHPsBSMSGNN} and Table \ref{Tab:tabHPsBSMSGNNTrain}, respectively. In our hyperparameter optimization on the 2D airfoil dataset, we observed the best results for the number of two neighbors per node, chosen as the nearest points to its left and right, in the input graphs. In the three-dimensional setting, the manual construction of a graph is less intuitive. Thus, on the NASA CRM dataset, we instead opted to use the adjacency matrix provided in the dataset itself, resulting in an input graph with a varying number of approximately four neighbors per node. Further, in the hyperparameter optimization in the 2D case the best results were attained for the number of $8$ scales. On the NASA CRM dataset we increased this number to $11$ in accordance with previous experiences with the model. As the seeding heuristic, determining the starting point of the bi-stride selection algorithm in each connected component of the graphs, we used the minimum average unweighted graph distance, corresponding to the model's default settings, in the 2D case. In this method, the seed node is chosen as the node with the shortest mean geodesic distance to all other nodes in the respective connected component. On the NASA CRM dataset, this method becomes infeasible due to the large amount of nodes. For this reason, in the 3D case we switched the seeding heuristic to the minimum distance to the center, choosing the seed node as the node with the lowest Euclidean distance to the center of the graph. Also, in accordance with the model's default settings, we chose to add self loops to the graphs, ensuring that all remaining nodes receive their own information in each downsampling step. In contrast to the model's default settings, we selected the aggregation method in the message passing mechanism as the mean function, since initial experimental training runs indicated this choice to be superior to the default summation function. Finally, on the 2D airfoil dataset, a batch size of $20$ lead to the best results during the hyperparameter optimization. However, as the model generally displayed a clear tendency of performing better for smaller batch sizes, in the 3D setting we reduced the batch size to $1$ in an attempt to improve the learning capability on the demanding NASA CRM dataset by allowing for more frequent parameter updates. We further adjusted the learning rate to $0.0002$, since this value resulted in the lowest average MAE ($0.011293$) in combination with the lowest batch size investigated in the hyperparameter optimization. An overview of the model's set up used in experiments carried out by its authors can be found in \cite[Appendix A]{derrickbsmsgnn}.

Transolver - which we evaluated on the 2D airfoil dataset - and Transolver++ - which we evaluated on the 3D NASA CRM dataset - largely coincide in their architectures. We thus combine our hyperparameter choices for these models in Table \ref{Tab:tabHPsTransolver} and Table \ref{Tab:tabHPsTransolverTrain}. We point out that, while both architectures share many hyperparameters, the published codes partially differ in their default and exemplary settings, which is the reason for most of the discrepancies between our choices regarding their respective set ups. The number of $2,500$ training epochs applied for Transolver on the 2D airfoil dataset was chosen as, in initial experimental runs, it had proved to be sufficient to achieve convergence to satisfactory results. For Transolver++ in the 3D setting, we increased the number of epochs to $10,000.$ In fact, in the 3D case this number was chosen to be the same for all models, allowing sufficient time for effective learning of the intricate details of the NASA CRM dataset. The batch size of $4$ chosen for Transolver++ on the NASA CRM dataset corresponds to the highest possible value not overstraining the model's capacities on an NVIDIA H100 GPU with 80GB HBM3. This choice was made to compensate for the high training time we experienced with Transolver++ on the NASA CRM dataset. Moreover, we notice that the learning rate scheduler used for the training process of both Transolver and Transolver++ corresponds to the PyTorch OneCycleLR scheduler. This scheduler comprises two phases: An initial growth period up to a specified maximum learning rate starting from a $25$ times smaller initial value, followed by a decay period during which the learning rate monotonously decreases to a fraction of the maximum learning rate specified by a final division factor. For a comparison to the set ups used by the models' authors across various experiments, we refer to \cite[Section B.3]{nnarticle} and \cite[Section B.3]{transolver++}, respectively.

Regarding the set up of the UPT architecture, we point out that UPT is specifically designed for evolutionary problems. This in particular shows in its processor, which is used to propagate the solution forward in time for the desired number of time steps. The architecture further includes a conditioner, consisting of a small MLP, for the time steps. As our experiments, however, are restricted to the steady state case, we drop the conditioner while employing the approximator for a single propagation step. Our hyperparameter choices concerning the architecture and the training process of the model are summarized in Table \ref{Tab:tabHPsUPT} and Table \ref{Tab:tabHPsUPTrain}, respectively. The value stated in Table \ref{Tab:tabHPsUPT} for the pooling radius in the 2D case corresponds to our considerations in Section \ref{radiuschoices}, where it was chosen in an attempt to achieve approximately 24 neighbors per supernode, cf.\ formula \eqref{radiusupt}. In the 3D case, we adjusted the pooling radius such that the neighborhood of each supernode would approximately cover the same portion of the physical domain as in the 2D setting, cf.\ formula \eqref{uptrad3d}. Moreover, while in the 2D setting our hyperparameter optimization led to a choice of $36$ supernodes and latent tokens, this number was not sufficient in the 3D case due to the sheer amount of datapoints in the NASA CRM dataset. Instead, in the latter case, we increased the number to $20480$ supernodes and $5120$ latent tokens. These numbers were chosen in accordance with the models' authors choices in \cite[Section 4.2]{upt}, where UPT was trained with $2048$ supernodes and $512$ latent tokens on a (transient flow) dataset of $29,000$--$59,000$ mesh points per sample. As the NASA CRM dataset contains $454,404$ and thus roughly ten times as many datapoints per sample, we decided to increase these numbers by the factor $10$. In the 3D setting, we furthermore set the number of training epochs to $10,000$ (in accordance with the set up of the remaining models) and the batch size to $1$ (in accordance with the set up of BSMS-GNN and GAOT). The decay rate of $0.0004$ lead to the best results during the hyperparameter optimization in both the cases of $10,000$ and $15,000$ training epochs, making it the adequate choice in both the 2D and the 3D setting. Eventually, we recall that for UPT we used a custom learning rate scheduler consisting of a warm up phase of linear growth during the first $10\%$ of epochs, followed by exponential decay for the remaining $90\%$ of epochs, which we warranted in the hyperparameter optimization in Section \ref{hpopt}.

Finally, the hyperparameter values chosen for GAOT are listed in Table \ref{Tab:tabHPsGAOT} and Table \ref{Tab:tabHPsGAOTTrain}, respectively. For the set up suggested by the models' authors we refer to \cite[Sections B.6.3, B.6.4]{gaot}. While the number of supernodes we used in the 2D case corresponds to the result of our hyperparameter optimization, we chose the supernodes in the 3D setting in accordance with these suggestions. We point out that in view of the rather high errors but low training time attained by GAOT on the NASA CRM dataset (cf.\ Tables \eqref{Tab:tabRes3d}, \eqref{Tab:tabRes3deff}), it might seem reasonable to attempt to increase the model's accuracy at the expense of the training speed by modifying more of its hyperparameters. However, we did not pursue that approach further after first attempts of increasing the pooling radius and the number of supernodes indicated to have the opposite effect. For the number of training epochs in the 2D setting, we relied, as in the case of Transolver, on our experience from initial experimental runs, indicating $5,000$ epochs to be adequate to reach convergence to adequate results.
In the 3D case, we again increased this number to $10,000$ as for the remaining models and chose the batch size as $1$ in accordance with our set ups of BSMS-GNN and UPT. We further point out that the learning rate scheduler referred to as \textit{Mix} in Table \ref{Tab:tabHPsGAOTTrain}, represents a custom scheduler set as default in the publicly available code. This scheduler splits the training of GAOT into three phases: During the first $2\%$ of epochs, the learning rate increases linearly from an initial to a maximum value. During the subsequent $96\%$ of epochs, the learning rate drops monotonously to a value referred to as the minimum learning rate following a cosine decay strategy. Ultimately, during the remaining $2\%$ of epochs, the learning rate decreases further to its final value following an exponential decay schedule.

\subsection{Evaluation metrics} \label{metrics}

We present precise mathematical definitions of the global error metrics used for the evaluation of the models in our experimental studies (cf.\ Section \ref{results2d} and \ref{resultsNASACRM}). These metrics consist of the mean absolute error (MAE), the mean squared error (MSE), the root mean squared error (RMSE), the relative $L^1$- and $L^2$-error (Rel. L1, Rel. L2) and the $R^2$-score (R2). We point out that the names of these error metrics here refer to how the error is calculated within individual samples, for each metric the final global value is then calculated as the mean of the errors on the individual samples. For example, Rel. L1 first takes the relative $L^1$ error over all spatial points within each data sample; subsequently it takes the mean of the resulting values across all data samples.

Let $n_s \in \mathbb{N}$ denote the number of data samples in a given dataset and let $n \in \mathbb{N}$ denote the number of spatial points per data sample in which a prediction is made. By $y^k_i \in \mathbb{R}$ we denote the prediction made by a model in the $i$-th spatial point of the $k$-th data sample, by $y^k := \{y_i^k\}_{i=1}^n$ we denote the set of predictions made on the $k$-th data sample and by $y := \{y^k\}_{k=1}^{n_s}$ we denote the set of predictions made on the entire dataset. Correspondingly, we denote by $\hat{y}_i^k$, $\hat{y}^k$ and $\hat{y}$ the associated ground truth values in individual spatial points, in data samples and on the entire dataset, respectively. Furthermore, by
\begin{align}
\overline{\hat{y}}^k := \frac{1}{n} \sum_{i=1}^n \hat{y}^k_i \nonumber
\end{align}

we denote the mean value of the ground truth values on the $k$-the data sample. With this notation at hand, we define the above error metrics in the following way:
\begin{align}
\text{MAE} \left( y, \hat{y} \right) :=& \frac{1}{n_s} \sum_{k=1}^{n_s} \left( \frac{1}{n} \sum_{i=1}^n \left| y_i^k - \hat{y}_i^k \right| \right), \nonumber \\
\text{MSE} \left( y, \hat{y} \right) :=& \frac{1}{n_s} \sum_{k=1}^{n_s} \left( \frac{1}{n} \sum_{i=1}^n \left| y_i^k - \hat{y}_i^k \right|^2 \right), \nonumber \\
\text{RMSE} \left( y, \hat{y} \right) :=& \frac{1}{n_s} \sum_{k=1}^{n_s} \sqrt{ \frac{1}{n} \sum_{i=1}^n \left| y_i^k - \hat{y}_i^k \right|^2}, \nonumber \\
\text{Rel. L1} \left( y, \hat{y} \right) :=& \frac{1}{n_s} \sum_{k=1}^{n_s} \left(\frac{\sum_{i=1}^n \left| y_i^k - \hat{y}_i^k \right|}{\sum_{i=1}^n \left| \hat{y}_i^k \right|}\right), \nonumber \\
\text{Rel. L2} \left( y, \hat{y} \right) :=& \frac{1}{n_s} \sum_{k=1}^{n_s} \left(\frac{\sqrt{\sum_{i=1}^n \left| y_i^k - \hat{y}_i^k \right|^2}}{\sqrt{\sum_{i=1}^n \left| \hat{y}_i^k \right|^2}}\right), \nonumber \\
\text{R2} \left( y, \hat{y} \right) :=& \frac{1}{n_s} \sum_{k=1}^{n_s} \left( 1 - \frac{\sum_{i=1}^n \left| \hat{y}_i^k - y_i^k \right|^2}{\sum_{i=1}^n \left| \hat{y}_i^k - \overline{\hat{y}}^k \right|^2}\right). \nonumber
\end{align}

\subsection{Supernode pooling radii} \label{radiuschoices}

The authors of the UPT model suggest to approximately choose the pooling radius $r > 0$ of the supernodes in the decoder of their architecture such that, on average, each supernode has $24$ neighbors in the set of input nodes, cf.\ \cite[Section F.1]{upt}. In our experiment on the 2D airfoil dataset in Section \ref{rae2822exp}, we tried to achieve this choice in the following way:

Let $N=512 \in \mathbb{N}$ denote the number of input nodes in each sample of the 2D airfoil dataset. Let $\gamma: [0, 1] \rightarrow \mathbb{R}^2$ denote a parametrization of the curve describing the 2D airfoil surface in an arbitrary data sample and let $L > 0$ denote its length. As the supernodes in UPT are randomly subsampled from the input nodes, we can fix an arbitrary supernode $\tilde{x}$ and identify it as $\tilde{x} = \gamma(s_0)$ for some $s_0 \in [0, 1]$. By
\begin{align}
A := A\left(s_0, r \right) := \left\lbrace \gamma(s) \in \mathbb{R}^2:\ s \in [0, 1],\ \left| \gamma(s) - \gamma\left(s_0 \right) \right| \leq r \right\rbrace \nonumber
\end{align}

\begin{figure}
\centering
\includegraphics[scale=0.5]{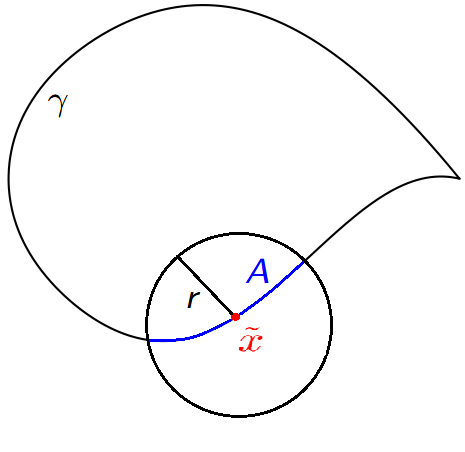} %[width=16cm,height=11cm]{solidinfluidnew}
%\hspace*{0.3cm}
\caption{The arc $A$ of the surface curve $\gamma$ lying within the pooling radius $r$ around the supernode $\tilde{x}$.} \label{figure1}
\end{figure}

we denote the arc of the curve $\gamma$ lying within the radius $r$ around $\tilde{x}$, cf.\ Figure \ref{figure1}. Assuming the original input nodes to be randomly sampled on the surface, each individual node has the probability
\begin{align}
p = \frac{|A|}{L}, \nonumber
\end{align}

of falling into $A$, where $|A|$ denotes the arc length of $A$. Thus, the number $K$ of remaining nodes (excluding $\tilde{x}$) in A satisfies
\begin{align}
\mathbb{E}[K] = (N-1)p = (N-1) \frac{|A|}{L} \approx (N-1) \frac{2r}{L}, \label{exp}
\end{align}

provided that $r$ is sufficiently small for $A$ to be close to a straight line, i.e.\ $|A| \approx 2r$. We point out that the latter assumption is reasonable due to the $C^2$-regularity of the curve $\gamma$, determined by a CST parametrization, cf.\ \cite[Section II.C]{rae2822}. Solving \eqref{exp} for $r$, we infer that
\begin{align}
r \approx \frac{L\mathbb{E}[K]}{2(N-1)}. \label{radius}
\end{align}

In the training set of the 2D airfoil dataset, the average length of the surface curves (after normalization via min-max scaling) amounts to $L \approx 2.63$. Consequently, under these assumptions, an average of $24$ neighbors per supernode in each (training) sample can be achieved by choosing 
\begin{align}
r = \frac{2.63 * 24}{2 * 511} \approx 0.062, \label{radiusupt}
\end{align}

which corresponds to our choice of the supernode pooling radius in the experimental set up for UPT in Section \ref{rae2822exp}. 

In our experiment on the NASA CRM dataset, we adjusted this number in the following way: Each neighborhood in the 2D setting, represented by a two-dimensional ball $B_r^2 \subset \mathbb{R}^2$ with radius $r = 0.062$, covers a portion of approximately
\begin{align}
\frac{\left| B_r^2 \right|}{1} = \frac{\pi r^2}{1} \approx 0.01207 \nonumber
\end{align}

of the (normalized) physical domain $[0,1] \times [0,1]$. We adjusted the pooling radius $r$ in such a way, that each three-dimensional neighborhood, represented by a three-dimensional ball $B_r^3 \subset \mathbb{R}^3$, would cover approximately the same portion of the corresponding (normalized) physical domain $[0, 1] \times [0, 1] \times [0, 1]$. More specifically, in the experimental set up for UPT on the NASA CRM dataset in Section \ref{nasacrm}, we chose the pooling radius $r$ according to
\begin{align}
\frac{\left| B_r^3 \right|}{1} = \frac{4\pi r^3}{3} \approx 0.01207 \quad \quad \Leftrightarrow \quad \quad r \approx 0.1423. \label{uptrad3d}
\end{align}

In the (publicly available version of the) GAOT model, the supernodes are not subsampled from the input nodes, but constructed as a regular grid in the entire physical space. For this reason, the supernodes do not lie on the airfoil surface and the pooling radius in case of the 2D airfoil dataset cannot be calculated in dependence of the desired number of neighbors per supernode by the same formula \eqref{radius} as for UPT. Instead, we decided to choose the pooling radius in such a way that 1.) the supernodes closest to the surface curve can be expected to possess at least the desired number of neighbors and 2.) each point on the surface curve lies within the pooling radius of a supernode. We attempted to achieve this in the following way:

In the experimental set up of our hyperparameter optimization, the (normalized) physical space is the unit square $\Omega := [0, 1] \times [0, 1] \subset \mathbb{R}^2$, in which $N_s \times N_s$, $N_s \in \{20, 30 \}$, supernodes are arranged as a regular grid with uniform distance $d = \frac{1}{19} \approx 0.053$ (if $N_s = 20$) or $d = \frac{1}{29} \approx 0.034$ (if $N_s = 30$) in both directions. Consequently, for each point $x = \gamma(s)$, $s \in [0, 1]$, on the surface curve $\gamma: [0, 1] \rightarrow \mathbb{R}^2$, there exists a supernode $\tilde{x} \in \Omega$ with distance
\begin{align}
\left| x - \tilde{x} \right| \leq \sqrt{\left( \frac{1}{2}d \right)^2 + \left( \frac{1}{2}d \right)^2 } = \sqrt{\frac{1}{2}}d =: \hat{d} \label{maxDist}
\end{align}

\begin{figure}
\centering
\includegraphics[scale=0.23]{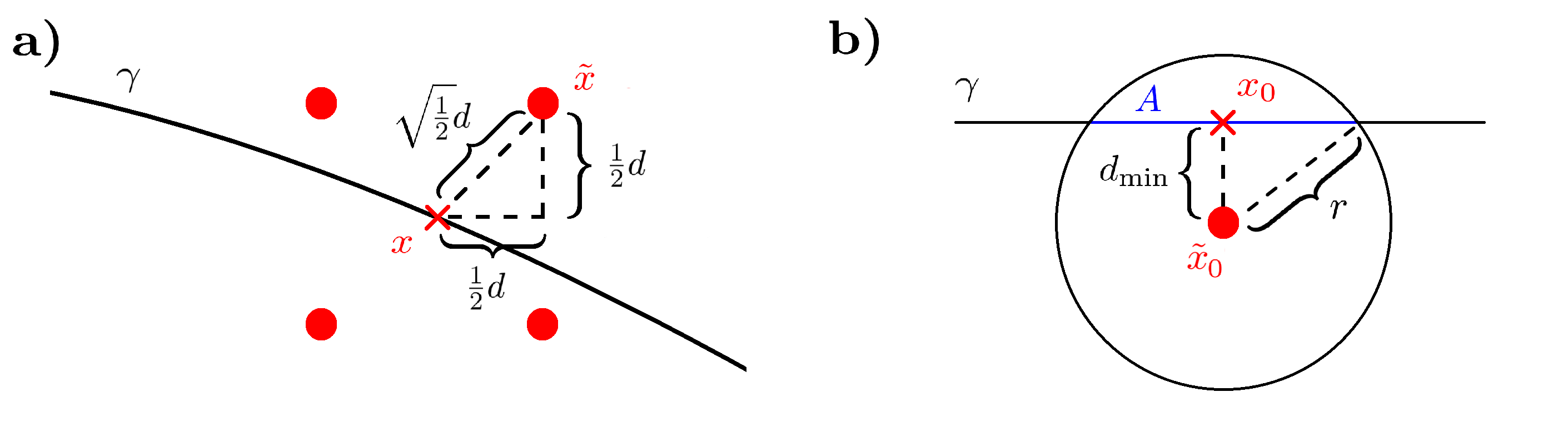} %[width=16cm,height=11cm]{solidinfluidnew}
%\hspace*{0.3cm}
\caption{\textbf{a)} Surface point $x$ with furthest possible distance $\hat{d} = \sqrt{\frac{1}{2}}d$ to a supernode $\tilde{x}$. \textbf{b)} The arc $A$ of the surface curve $\gamma$ lying within the pooling radius $r$ around the supernode $\tilde{x}_0$.} \label{figure2}
\end{figure}

to the surface point $x$, cf.\ Figure \ref{figure2} a). Let
\begin{align}
S := \left\lbrace \tilde{x} \in \Omega:\ \tilde{x} \text{ is supernode, } \left| \tilde{x} - \gamma(s) \right| \leq \hat{d}\ \text{for some } s \in [0,1] \right\rbrace \nonumber
\end{align}

denote the set of all supernodes lying within the distance $\hat{d}$ of the surface curve and fix an arbitrary supernode $\tilde{x}_0 \in S$. Let $x_0 = \gamma(s_0)$, $s_0 \in [0, 1]$, denote the surface point with the shortest distance
\begin{align}
d_{\min} := \min \left\lbrace \left| \tilde{x}_0 - \gamma(s) \right|:\ s \in [0, 1] \right\rbrace \leq \hat{d}. \nonumber
\end{align}

to $\tilde{x}_0$. We denote by
\begin{align}
A := A\left(\tilde{x}_0, r \right) := \left\lbrace \gamma(s) \in \Omega:\ s \in [0, 1],\ \left| \tilde{x}_0 - \gamma(s) \right| \leq r \right\rbrace \nonumber
\end{align}

the arc of the curve $\gamma$ lying within the pooling radius $r>0$ of the supernode $\tilde{x}_0$. Again, we assume $r$ to be sufficiently small for $A$ to be approximately a straight line. Since the straight line connecting the supernode $\tilde{x}_0$ to the surface point $x_0$ is perpendicular to $\gamma$, it follows that
\begin{align}
|A| \approx 2\sqrt{r^2 - d_{\min}^2} \geq 2\sqrt{r^2 - \hat{d}^2} = 2\sqrt{r^2 - \frac{1}{2} d^2}, \nonumber
\end{align}

cf.\ Figure \ref{figure2} b). Similar to the formula \eqref{exp}, it holds that the number $K$ of input nodes falling into $A$ obeys
\begin{align}
\mathbb{E}[K] = N \frac{|A|}{L} \gtrsim 2N \frac{\sqrt{r^2 - \frac{1}{2} d^2}}{L} \quad \quad \Leftrightarrow \quad \quad \sqrt{\left( \frac{L \mathbb{E}[K] }{2N} \right)^2 + \frac{1}{2}d^2} \gtrsim r, \nonumber
\end{align}

where again $N=512 \in \mathbb{N}$ denotes the number of input nodes per sample in the 2D airfoil dataset and $L > 0$ represents the length of $\gamma$. Plugging the average length $L\approx 2.63$ of the normalized surface curves in the training set, the number $N=512$ of input nodes as well as the step size $d \in \{ \frac{1}{19}, \frac{1}{29} \}$ between the supernodes into this formula, we can achieve the desired minimum expected number of neighbors per supernode in the set $S$ by choosing the radius $r$ sufficiently large. We further point out that, since each input node lies within the $\hat{d}$-radius of a supernode (cf.\ estimate \eqref{maxDist}) and since, by construction, $r \geq \hat{d}$, each input node also lies within the neighborhood of at least one supernode. In order to ensure at least 24 expected neighbors per supernode in $S$, in compliance with the set up of GAOT in the hyperparameter optimization in Section \ref{hpopt}, we set
\begin{align}
r = 0.072 \approx \sqrt{\left( \frac{2.63 * 24 }{2 * 512} \right)^2 + \frac{1}{2} * \left(\frac{1}{19}\right)^2} \quad &\text{if } N_s = 20\ \left(\text{i.e. } d = \frac{1}{19}\right), \nonumber \\
r = 0.066 \approx \sqrt{\left( \frac{2.63 * 24 }{2 * 512} \right)^2 + \frac{1}{2} * \left(\frac{1}{29}\right)^2} \quad &\text{if } N_s = 30\ \left(\text{i.e. } d = \frac{1}{29}\right). \nonumber
\end{align}

\bibliographystyle{cas-model2-names}

\bibliography{cas-refs}

@inproceedings{upt,
 author = {Alkin, B. and Fürst, A. and Schmid, S. and Gruber, L. and Holzleitner, M. and Brandstetter, J.},
 booktitle = {Advances in Neural Information Processing Systems},
 doi = {10.52202/079017-0793},
 pages = {25152--25194},
 title = {Universal Physics Transformers: A Framework For Efficiently Scaling Neural Operators},
 volume = {37},
 year = {2024}
}

@misc{drivaerml,
  author = {Ashton, N. and Mockett, C. and Fuchs, M. and Fliessbach, L. and Hetmann, H. and Knacke, T. and Schonwald, N. and Skaperdas, V. and Fotiadis, G. and Walle, A. and Hupertz, B. and Maddix, D.},
  title  = {DrivAerML: High-Fidelity Computational Fluid Dynamics Dataset for Road-Car External Aerodynamics},
  howpublished = "\url{https://doi.org/10.48550/arXiv.2408.11969}",
  year = {2024}
}

@inproceedings{rae2822,
 author = {Bekemeyer, P. and Hariharan, N. and Wissink, A.M. and Cornelius, J.},
 booktitle = {AIAA SCITECH 2025 Forum},
 doi = {10.2514/6.2025-0036},
 title = {Introduction of Applied Aerodynamics Surrogate Modeling Benchmark Cases},
 year = {2025}
}

@misc{abupt,
  author  = {Bleeker, M.J.R. and Dorfer, M. and Kronlachner, T. and Sonnleitner, R. and Alkin, B. and Brandstetter, J.},
  title   = {AB-UPT: Scaling Neural CFD Surrogates for High-Fidelity Automotive Aerodynamics Simulations via Anchored-Branched Universal Physics Transformers},
  howpublished = "\url{https://doi.org/10.48550/arXiv.2502.09692}",
  year    = {2025}
}

@inproceedings{airfrans,
 author = {Bonnet, F. and Mazari, J. and Cinnella, P. and Gallinari, P.},
 booktitle = {Advances in Neural Information Processing Systems},
 doi = {10.52202/068431-1705},
 pages = {23463--23478},
 title = {AirfRANS: High Fidelity Computational Fluid Dynamics Dataset for Approximating Reynolds-Averaged Navier\textendash Stokes Solutions},
 volume = {35},
 year = {2022}
}

@inproceedings{gnn1,
  title     = {Message Passing Neural PDE Solvers},
  author    = {Brandstetter, J. and Worrall, D.E. and Welling, M.},
  booktitle = {International Conference on Learning Representations},
  year      = {2022},
  url       = {https://mlanthology.org/iclr/2022/brandstetter2022iclr-message/}
}

@inproceedings{chooseatransformer,
 author = {Cao, S.},
 booktitle = {Advances in Neural Information Processing Systems},
 pages = {24924--24940},
 title = {Choose a Transformer: Fourier or Galerkin},
 url = {https://proceedings.neurips.cc/paper_files/paper/2021/file/d0921d442ee91b896ad95059d13df618-Paper.pdf},
 volume = {34},
 year = {2021}
}

@InProceedings{bsmsgnn,
  title =  {Efficient Learning of Mesh-Based Physical Simulation with Bi-Stride Multi-Scale Graph Neural Network},
  author =  {Cao, Y. and Chai, M. and Li, M. and Jiang, C.},
  booktitle =  {Proceedings of the 40th International Conference on Machine Learning},
  pages =  {3541--3558},
  year =  {2023},
  volume =  {202},
  url =  {https://proceedings.mlr.press/v202/cao23a.html},
}

@inproceedings{elrefaie,
 author = {Elrefaie, M. and Ayman, T. and Elrefaie, M. and Sayed, E. and Ayyad, M. and AbdelRahman, M.M.},
 booktitle = {AIAA SCITECH 2024 Forum},
 doi = {10.2514/6.2024-2220},
 title = {Surrogate Modeling of the Aerodynamic Performance for Airfoils in Transonic Regime},
 year = {2024}
}

@article{drivaernet,
    author = {Elrefaie, M. and Dai, A. and Ahmed, F.},
    title = {DrivAerNet: A Parametric Car Dataset for Data-Driven Aerodynamic Design and Prediction},
    journal = {Journal of Mechanical Design},
    volume = {147},
    number = {4},
    pages = {041712},
    year = {2025},
    month = {03},
    doi = {10.1115/1.4068104}
}

@inproceedings{drivaernet++,
 author = {Elrefaie, M. and Morar, F. and Dai, A. and Ahmed, F.},
 booktitle = {Advances in Neural Information Processing Systems},
 doi = {10.52202/079017-0016},
 pages = {499--536},
 title = {DrivAerNet++: A Large-Scale Multimodal Car Dataset with Computational Fluid Dynamics Simulations and Deep Learning Benchmarks},
 volume = {37},
 year = {2024}
}

@misc{carbench,
  author = {Elrefaie, M. and Shu, D. and Klenk, M. and Ahmed, F.},
  title  = {CarBench: A Comprehensive Benchmark for Neural Surrogates on High-Fidelity 3D Car Aerodynamics},
  howpublished = "\url{https://doi.org/10.48550/arXiv.2512.07847}",
  year = {2025}
}

@book{evans,
  title     = "Partial Differential Equations",
  author    = "Evans, L.C.",
  year      = 1998,
  publisher = "American Mathematical Society",
  address   = "Providence, Rhode Island"
}

@InProceedings{gnot,
  title =  {{GNOT}: A General Neural Operator Transformer for Operator Learning},
  author = {Hao, Zhongkai and Wang, Zhengyi and Su, Hang and Ying, Chengyang and Dong, Yinpeng and Liu, Songming and Cheng, Ze and Song, Jian and Zhu, Jun},     
  booktitle =  {Proceedings of the 40th International Conference on Machine Learning},
  pages =  {12556--12569},
  year =  {2023},
  volume =  {202},
  url =  {https://proceedings.mlr.press/v202/hao23c.html},
}

@inproceedings{poseidon,
 author = {Herde, Maximilian and Raoni\'{c}, Bogdan and Rohner, Tobias and K\"{a}ppeli, Roger and Molinaro, Roberto and de B\'{e}zenac, Emmanuel and Mishra, Siddhartha},
 booktitle = {Advances in Neural Information Processing Systems},
 doi = {10.52202/079017-2311},
 pages = {72525--72624},
 title = {Poseidon: Efficient Foundation Models for PDEs},
 volume = {37},
 year = {2024}
}

@article{derrickgns,
title = {Graph neural networks for the prediction of aircraft surface pressure distributions},
journal = {Aerospace Science and Technology},
volume = {137},
pages = {108268},
year = {2023},
issn = {1270-9638},
doi = {https://doi.org/10.1016/j.ast.2023.108268},
author = {Derrick Hines and Philipp Bekemeyer},
}

@article{derrickbsmsgnn,
title = {Prediction of surface pressure distributions of non-parametric airfoils using geometric deep learning methods},
journal = {Computers \& Fluids},
volume = {308},
pages = {106979},
year = {2026},
doi = {https://doi.org/10.1016/j.compfluid.2026.106979},
author = {Derrick Hines and Philipp Bekemeyer},
}

@article{noarticle,
  author  = {Nikola Kovachki and Zongyi Li and Burigede Liu and Kamyar Azizzadenesheli and Kaushik Bhattacharya and Andrew Stuart and Anima Anandkumar},
  title   = {Neural Operator: Learning Maps Between Function Spaces With Applications to PDEs},
  journal = {Journal of Machine Learning Research},
  year    = {2023},
  volume  = {24},
  number  = {89},
  pages   = {1--97},
  url     = {http://jmlr.org/papers/v24/21-1524.html}
}

@article{geofno,
  author  = {Zongyi Li and Daniel Zhengyu Huang and Burigede Liu and Anima Anandkumar},
  title   = {Fourier Neural Operator with Learned Deformations for PDEs on General Geometries},
  journal = {Journal of Machine Learning Research},
  year    = {2023},
  volume  = {24},
  number  = {388},
  pages   = {1--26},
  url     = {http://jmlr.org/papers/v24/23-0064.html}
}

@misc{oformer,
  author = {Li, Z. and Meidani, K. and Farimani, A.},
  title  = {Transformer for Partial Differential Equations' Operator Learning},
  howpublished = "\url{https://doi.org/10.48550/arXiv.2205.13671}",
  year = {2022}
}

@inproceedings{mgno,
 author = {Li, Zongyi and Kovachki, Nikola and Azizzadenesheli, Kamyar and Liu, Burigede and Stuart, Andrew and Bhattacharya, Kaushik and Anandkumar, Anima},
 booktitle = {Advances in Neural Information Processing Systems},
 pages = {6755--6766},
 title = {Multipole Graph Neural Operator for Parametric Partial Differential Equations},
 url = {https://proceedings.neurips.cc/paper_files/paper/2020/file/4b21cf96d4cf612f239a6c322b10c8fe-Paper.pdf},
 volume = {33},
 year = {2020}
}

@misc{fno,
  author = {Li, Z. and Kovachki, N. and Azizzadenesheli, K. and Liu, B. and Bhattacharya, K. and Stuart, A. and Anandkumar, A.},
  title  = {Fourier Neural Operator for Parametric Partial Differential Equations},
  howpublished = "\url{https://doi.org/10.48550/arXiv.2010.08895}",
  year = {2020}
}

@misc{gno,
  author = {Li, Z. and Kovachki, N. and Azizzadenesheli, K. and Liu, B. and Bhattacharya, K. and Stuart, A. and Anandkumar, A.},
  title  = {Neural Operator: Graph Kernel Network for Partial Differential Equations},
  howpublished = "\url{https://doi.org/10.48550/arXiv.2003.03485}",
  year = {2020}
}

@inproceedings{gino,
 author = {Li, Zongyi and Kovachki, Nikola and Choy, Chris and Li, Boyi and Kossaifi, Jean and Otta, Shourya and Nabian, Mohammad Amin and Stadler, Maximilian and Hundt, Christian and Azizzadenesheli, Kamyar and Anandkumar, Animashree},
 booktitle = {Advances in Neural Information Processing Systems},
 pages = {35836--35854},
 title = {Geometry-Informed Neural Operator for Large-Scale 3D PDEs},
 doi = {10.52202/075280-1556},
 volume = {36},
 year = {2023}
}

@article{pino,
author = {Li, Zongyi and Zheng, Hongkai and Kovachki, Nikola and Jin, David and Chen, Haoxuan and Liu, Burigede and Azizzadenesheli, Kamyar and Anandkumar, Anima},
title = {Physics-Informed Neural Operator for Learning Partial Differential Equations},
year = {2024},
volume = {1},
number = {3},
journal = {ACM / IMS J. Data Sci.},
doi = {10.1145/3648506},
pages = {9},
numpages = {27},
}

@article{nosurvey,
title = {Architectures, variants, and performance of neural operators: A comparative review},
journal = {Neurocomputing},
volume = {648},
pages = {130518},
year = {2025},
doi = {https://doi.org/10.1016/j.neucom.2025.130518},
author = {Shengjun Liu and Yu Yu and Ting Zhang and Hanchao Liu and Xinru Liu and Deyu Meng},
}

@misc{deeponet,
  author = {Lu, L. and Jin, P. and Karniadakis, G.E.},
  title  = {Deeponet: Learning nonlinear operators for identifying differential equations based on the universal approximation theorem of operators},
  howpublished = "\url{https://doi.org/10.1038/s42256-021-00302-5}",
  year = {2019}
}

@misc{transolver++,
  author = {Luo, H. and Wu, H. and Zhou, H. and Xing, L. and Di, Y. and Wang, J. and Long, M.},
  title  = {Transolver++: An Accurate Neural Solver for PDEs on Million-Scale Geometries},
  howpublished = "\url{https://doi.org/10.48550/arXiv.2502.02414}",
  year = {2025}
}

@inproceedings{rigno,
 author = {Mousavi, Sepehr and Wen, Shizheng and Lingsch, Levi and Herde, Maximilian and Raonic, Bogdan and Mishra, Siddhartha},
 booktitle = {Advances in Neural Information Processing Systems},
 pages = {150039--150101},
 title = {RIGNO: A Graph-based Framework For Robust And Accurate Operator Learning For PDEs On Arbitrary Domains},
 url = {https://proceedings.neurips.cc/paper_files/paper/2025/file/dcb91f43033bb1d367d1848806dee98d-Paper-Conference.pdf},
 volume = {38},
 year = {2025}
}

@misc{gnn2,
  author = {Pfaff, T. and Fortunato, M. and Sanchez-Gonzalez, A. and Battaglia, P.},
  title  = {Learning mesh-based simulation with graph networks},
  howpublished = "\url{https://doi.org/10.48550/arXiv.2010.03409}",
  year = {2020}
}

@book{numerics,
  title     = "Numerical Approximation of Partial Differential Equations",
  author    = "Quarteroni, A. and Valli, A.",
  year      = 1994,
  publisher = "Springer Verlag",
  address   = "Berlin, Heidelberg"
}

@article{pinns,
title = {Physics-informed neural networks: A deep learning framework for solving forward and inverse problems involving nonlinear partial differential equations},
journal = {Journal of Computational Physics},
volume = {378},
pages = {686-707},
year = {2019},
doi = {https://doi.org/10.1016/j.jcp.2018.10.045},
author = {M. Raissi and P. Perdikaris and G.E. Karniadakis},
}

@misc{domino,
  author = {Ranade, R. and Nabian, M.A. and Tangsali, K. and Kamenev, A. and Hennigh, O. and Cherukuri R. and Choudhry, S.},
  title  = {DoMINO: A Decomposable Multi-scale Iterative Neural Operator for Modeling Large Scale Engineering Simulations},
  howpublished = "\url{https://doi.org/10.48550/arXiv.2501.13350}",
  year = {2025}
}

@inproceedings{cno,
 author = {Raonic, Bogdan and Molinaro, Roberto and De Ryck, Tim and Rohner, Tobias and Bartolucci, Francesca and Alaifari, Rima and Mishra, Siddhartha and de B\'{e}zenac, Emmanuel},
 booktitle = {Advances in Neural Information Processing Systems},
 pages = {77187--77200},
 title = {Convolutional Neural Operators for robust and accurate learning of PDEs},
 doi = {10.52202/075280-3376},
 volume = {36},
 year = {2023}
}

@book{roubicek,
  title     = "Nonlinear Partial Differential Equations with Applications",
  author    = "Roubí\v{c}ek, T.",
  year      = 2005,
  publisher = "Birkhäuser Verlag",
  address   = "Basel, Boston, Berlin"
}

@inproceedings{blendnet,
    author = {Sung, Nicholas and Spreizer, Steven and Elrefaie, Mohamed and Samuel, Kaira and Jones, Matthew C. and Ahmed, Faez},
    booktitle = {ASME 2025 International Design Engineering Technical Conferences and Computers and Information in Engineering Conference},
    title = {BlendedNet: A Blended Wing Body Aircraft Dataset and Surrogate Model for Aerodynamic Predictions},
    volume = {Volume 3B: 51st Design Automation Conference (DAC)},
    series = {International Design Engineering Technical Conferences and Computers and Information in Engineering Conference},
    pages = {V03BT03A049},
    year = {2025},
    doi = {10.1115/DETC2025-168977}
}

@inproceedings{transformers,
 author = {Vaswani, Ashish and Shazeer, Noam and Parmar, Niki and Uszkoreit, Jakob and Jones, Llion and Gomez, Aidan N and Kaiser, Lukasz and Polosukhin, Illia},
 booktitle = {Advances in Neural Information Processing Systems},
 title = {Attention is All you Need},
 url = {https://proceedings.neurips.cc/paper_files/paper/2017/file/3f5ee243547dee91fbd053c1c4a845aa-Paper.pdf},
 pages = {6000–6010},
 volume = {30},
 year = {2017}
}

@misc{gaot,
  author = {Wen, S. and Kumbhat, A. and Lingsch, L. and Mousavi, S. and Zhao, Y. and Chandrashekar, P. and Mishra, S.},
  title  = {Geometry Aware Operator Transformer as an Efficient and Accurate Neural Surrogate for PDEs on Arbitrary Domains},
  howpublished = "\url{https://doi.org/10.48550/arXiv.2505.18781}",
  year = {2025}
}

@misc{nnarticle,
  author = {Wu, H. and Luo, H. and Wang, H. and Wang, J. and Long, M.},
  title  = {Transolver: A Fast Transformer Solver for PDEs on General Geometries},
  howpublished = "\url{https://doi.org/10.48550/arXiv.2402.02366}",
  year = {2024}
}

@misc{transolver3,
  author = {Zhou, H. and Wu, H. and Shangguan, H. and Ma, Y. and Weng, H. and Wang, J. and Long, M.},
  title  = {Transolver-3: Scaling Up Transformer Solvers to Industrial-Scale Geometries},
  howpublished = "\url{https://doi.org/10.48550/arXiv.2602.04940}",
  year = {2026}
}



\clearpage

\begin{table}[]
\caption{Input and output features used in the study on the 2D airfoil dataset.} \label{Tab:tab1}
\begin{tabular}{c c c}
\toprule
\textbf{Input features} & Lower bound & Upper bound \\ [0.5ex]
\midrule
Spatial coordinates $x$ & - & - \\
% \hline
Surface normals n$_x$ & - & - \\
% \hline
Mach number $M$ & $0.2$ & $0.7$ \\
% \hline
Reynolds number $Re$ & $1.0 * 10^6$ & $6.0*10^6$ \\
% \hline
Angle of attak $AoA$ & $-3.0$ & $5.0$ \\
\midrule
\textbf{Output features} & Lower bound & Upper bound \\ [0.5ex]
\midrule
Pressure coefficient $C_p$ & -6.41 & 1.15 \\
\bottomrule
\end{tabular}
\end{table}

\begin{table}[]
\caption{Evaluated hyperparameter combinations for BSMS-GNN. The choices leading to the best result marked in bold.} \label{Tab:tab4}
\begin{tabular}{c  c} 
\toprule
Hyperparameters & Values \\ [0.5ex] 
\midrule
Max. learning rate & 0.0002, \textbf{0.00084}, 0.002 \\
% \hline
Batch size & 5, 10, \textbf{20} \\
% \hline
Neighbors per node & \textbf{2}, 8 \\
% \hline
Number of scales & 4, \textbf{8} \\
\bottomrule
\end{tabular}
\end{table}

\begin{table}[]
\caption{Evaluated hyperparameter combinations for Transolver. The choices leading to the best result marked in bold.} \label{Tab:tab-1}
\begin{tabular}{c  c} 
\toprule
Hyperparameters & Values \\ [0.5ex] 
\midrule
Max. learning rate & \textbf{0.0001}, 0.0002, 0.001 \\
% \hline
Batch size & \textbf{8}, 16, 32 \\
% \hline
Number of slices & 32, \textbf{64}, 128 \\ 
\bottomrule
\end{tabular}
\end{table}

\begin{table}[]
\caption{Evaluated hyperparameter combinations during the first and second round of hyperparameter optimization for UPT. In the first round, a learning rate scheduler with linear decay was employed; in the second round exponential decay was applied. Choices leading to the best results, respectively, marked in bold.} \label{Tab:tab3}
\begin{tabular}{c c}
\toprule
Hyperparameters (1st round) & Values \\ [0.5ex]
\midrule
Max. learning rate & 0.0001, \textbf{0.0002}, 0.001 \\
% \hline
Batch size & \textbf{8}, 16, 32 \\
% \hline
(Supernodes, Latent tokens) & (18, 4), (18, 18), (36, 9), \textbf{(36, 36)}, (72, 18), (72, 72) \\
\midrule
Hyperparameters (2nd round) & Values \\ [0.5ex]
\midrule
Max. learning rate & 0.0001, \textbf{0.0002}, 0.001 \\
% \hline
(Epochs, Decay rate) & (5000, 0.00080), (5000, 0.00120), (10000, 0.00040), \\
& (10000, 0.00060), (15000, 0.0002$\overline{6}$), \textbf{(15000, 0.00040)} \\
\bottomrule
\end{tabular}
\end{table}

\begin{table}[]
\caption{Evaluated hyperparameter combinations for GAOT. The choices leading to the best result marked in bold.} \label{Tab:tab2}
\begin{tabular}{c c} 
\toprule
Hyperparameters & Values \\ [0.5ex] 
\midrule
Max. learning rate & 0.0001, 0.0002, \textbf{0.001} \\
% \hline
Batch size & \textbf{8}, 16, 32 \\
% \hline
(Supernodes, Pooling radius) & (400, 0.038), (400, 0.072), (900, 0.026), \textbf{(900, 0.066)} \\ 
\bottomrule
\end{tabular}
\end{table}

\begin{table}[]
\caption{$95\%$ confidence intervals resulting from the evaluation of ten fully trained instantiations of each model on the test samples of the 2D airfoil dataset. The models are evaluated in terms of mean absolute error (MAE), mean squared error (MSE), root mean squared error (RMSE), relative $L^1$-error (Rel. L1), relative $L^2$-error (Rel. L2) and $R^2$-score (R2). A precise definition of the individual metrics is given in Appendix \ref{metrics}.} \label{Tab:tabRes2d}
\begin{tabular}{c c c c c}
\toprule
& BSMS-GNN & Transolver & UPT & GAOT \\ [0.5ex]
\midrule
MAE & $0.007178 \pm 0.000154$ & $0.006525 \pm 0.000190$ & $0.012383 \pm 0.000484$ & $0.008311 \pm 0.000174$ \\
% \hline
MSE & $0.001131 \pm 0.000071$ & $0.001018 \pm 0.000053$ & $0.001800 \pm 0.000153$ & $0.001392 \pm 0.000111$ \\
% \hline
RMSE & $0.014579 \pm 0.000231$ & $0.013225 \pm 0.000355$ & $0.020773 \pm 0.000633$ & $0.016572 \pm 0.000439$ \\
% \hline
Rel. L1 & $0.016895 \pm 0.000340$ & $0.015102 \pm 0.000384$ & $0.030047 \pm 0.001109$ & $0.019731 \pm 0.000365$ \\
% \hline
Rel. L2 & $0.025529 \pm 0.000405$ & $0.022476 \pm 0.000564$ & $0.037708 \pm 0.001093$ & $0.029123 \pm 0.000680$ \\
% \hline
R2 & $0.998730 \pm 0.000082$ & $0.998863 \pm 0.000067$ & $0.997897 \pm 0.000163$ & $0.998433 \pm 0.000110$ \\
\bottomrule
\end{tabular}
\end{table}

\begin{table}[]
\caption{Computational effiency of each model on the 2D airfoil dataset. All models were trained and evaluated on an NVIDIA Quadro P2200 GPU with 5GB memory. The number of parameters is measured for instantiations of the models with optimized hyperparameters according to Section \ref{hpopt}. The number of training epochs corresponds to the number of epochs used in the training of these instantiations leading up to the results in Table \ref{Tab:tabRes2d} and the average time per training epoch measures the time in seconds each of the models required for a single training epoch on the 498 training samples. The wall-clock time is calculated as the product of the latter two values and is stated in the format hours - minutes - seconds. The inference latency is measured as the average duration in milliseconds of a forward pass on a single sample in evaluation mode.} \label{Tab:tabRes2deff}
\begin{tabular}{c c c c c} 
\toprule
& BSMS-GNN & Transolver & UPT & GAOT \\ [0.5ex]
\midrule
Number of parameters & $426,113$ & $714,945$ & $1,654,177$ & $3,396,289$ \\
Number of training epochs & $10,000$ & $2,500$ & $15,000$ & $5,000$ \\
Avg. time per training epoch & $3.51$s & $3.91$s & $2.43$s & $5.33$s \\
Wall-clock time & $9$h $44$m $15$s &$2$h $42$m $43$s & $10$h $8$m $30$s & $7$h $24$m $1$s \\
Inference latency & $\leq 15$ms & $\leq 10$ms & $\leq 10$ms & $\leq 10$ms \\
\bottomrule
\end{tabular}
\end{table}

\begin{table}[]
\caption{Input and output features used in the study on the NASA CRM dataset.} \label{Tab:tabCRM}
\begin{tabular}{c c c} 
\toprule
\textbf{Input features} & Lower bound & Upper bound \\ [0.5ex]
\midrule
Spatial coordinates $x$ & - & - \\
Surface normals n$_x$ & - & - \\
Inboard aileron angle & $-20.0$ & $20.0$ \\
Outboard aileron angle & $-20.0$ & $10.0$ \\
Elevator angle & $-10.0$ & $10.0$ \\
HTP angle & $-2.0$ & $2.0$ \\
Mach number $M$ & $0.5$ & $0.88$ \\
Angle of attak $AoA$ & $-2.5$ & $5.0$ \\
\midrule
\textbf{Output features} & Lower bound & Upper bound \\ [0.5ex]
\midrule
Pressure coefficient $C_p$ & -3.0 & 2.27 \\
\bottomrule
\end{tabular}
\end{table}

\begin{table}[]
\caption{Evaluation of one fully trained instance of each architecture on the NASA CRM test set. The models are evaluated in terms of the global metrics defined in Appendix \ref{metrics}.} \label{Tab:tabRes3d}
\begin{tabular}{c c c c c}
\toprule
& BSMS-GNN & Transolver++ & UPT & GAOT \\ [0.5ex]
\midrule
MAE & $0.008298$ & $0.011092$ & $0.054048$ & $0.023976$ \\
MSE & $0.000651$ & $0.001034$ & $0.007295$ & $0.002444$ \\
RMSE & $0.019251$ & $0.025366$ & $0.082765$ & $0.046269$ \\
Rel. L1 & $0.035710$ & $0.047499$ & $0.240316$ & $0.105491$ \\
Rel. L2 & $0.057971$ & $0.076584$ & $0.264838$ & $0.146275$ \\
R2 & $0.996884$ & $0.995102$ & $0.957038$ & $0.986390$ \\
\bottomrule
\end{tabular}
\end{table}

\begin{table}[]
\caption{Computational effiency of the models' instantiations on the NASA CRM dataset. All models were trained and evaluated on an NVIDIA H100 GPU with 80GB HBM3.} \label{Tab:tabRes3deff}
\begin{tabular}{c c c c c} 
\toprule
& BSMS-GNN & Transolver++ & UPT & GAOT \\ [0.5ex]
\midrule
Number of parameters & $578,241$ & $1,751,589$ & $2,142,625$ & $3,856,881$ \\
Number of training epochs & $10,000$ & $10,000$ & $10,000$ & $10,000$ \\
Avg. time per training epoch & $37.30$s & $58.38$s & $55.88$s & $37.73$s \\
Wall-clock time & $4$d $7$h $36$m $40$s & $6$d $18$h $10$m $0$s & $6$d $11$h $13$m $20$s & $4$d $8$h $48$m $20$s \\
Inference latency & $\leq 95$ms & $\leq 165$ms & $\leq 260$ms & $\leq 110$ms \\
\bottomrule
\end{tabular}
\end{table}

\begin{table}[]
\caption{Hyperparameter values chosen in the set up of BSMS-GNN for surface pressure predictions on 2D airfoils in Section \ref{rae2822exp} and the 3D NASA CRM aircraft configuration in Section \ref{nasacrm}, respectively. Optimized hyperparameters on the 2D airfoil dataset marked in bold.} \label{Tab:tabHPsBSMSGNN}
\begin{tabular}{p{3.5cm} p{2cm} p{2cm} p{7cm}}
\toprule
Hyperparameter & Value (2D) & Value (3D) & Description \\[0.5ex]
\midrule
\textbf{Neighbors} & 2 & Varying & Number of neighbors per node in the input graph. \\
\textbf{Scales} & $8$ & $11$ & Number of scales/ coarseness levels, determining the depth of the GNN processor. \\
Cycles Processor & $1$ & $1$ & Number of times the processor is applied. \\
Seeding heuristic & Min.\ aver\-age graph distance & Min.\ distance to center & Method to determine the seed nodes in the bi-stride algorithm. \\
Aggr. method & mean & mean & Aggregation function for the message passing mechanism. \\
Self Loops & True & True & Wether to connect each node to itself in all graphs. \\
Complete Edges & True & True & Wether to connect all nodes on the coarsest scale to ensure exchange between different connected components. \\
Update positions & True & True & Wether to update the nodes' positions in each downsampling step. \\
Latent dim. & 64 & 64 & In- and output dimension of the GNN processor. \\
Message passing layers & 1 & 1 & Number of message passing steps on each scale. \\
Processor MLP layers & 1 & 1 & Number of layers in each MLP in the GNN processor. \\
Processor MLP neurons & 64 & 64 & Number of neurons per hidden layer in each MLP in the GNN processor. \\
Encoder MLP layers & 2 & 2 & Number of layers in the encoder MLP. \\
Encoder MLP neurons & 128 & 128 & Number of neurons per hidden layer in the encoder MLP. \\
Decoder MLP layers & 2 & 2 & Number of layers in the decoder MLP. \\
Decoder MLP neurons & 128 & 128 & Number of neurons per hidden layer in the decoder MLP. \\
Activation & elu & elu & Activation function used in all MLPs. \\
Dropout & 0.0 & 0.0 & Dropout probability in all MLPs. \\
\bottomrule
\end{tabular}
\end{table}

\begin{table}[]
\caption{Hyperparameter choices in the training procedure of BSMS-GNN for surface pressure predictions on 2D airfoils in Section \ref{rae2822exp} and the 3D NASA CRM aircraft configuration in Section \ref{nasacrm}, respectively. Optimized hyperparameters on the 2D airfoil dataset marked in bold.} \label{Tab:tabHPsBSMSGNNTrain}
\begin{tabular}{p{3.5cm} p{2cm} p{2cm} p{7cm}}
\toprule
Hyperparameter & Value (2D) & Value (3D) & Description \\ [0.5ex]
\midrule
Training epochs & $10,000$ & $10,000$ & Number of training epochs. \\
\textbf{Batch size} & $20$ & $1$ & Size of the training batches. \\
Loss function & MSE & MSE & Mean squared error loss function used in the training process. \\
Optimizer & Adam & Adam & Optimization algorithm. \\
Weight decay & $0.0$ & $0.0$ & Weight decay coefficient. \\
Learning rate sched. & Exp. decay & Exp. decay & Exponential decay from a max. initial value to a min. final value. \\
\textbf{Max. learning rate} & $0.00084$ & $0.0002$ & Maximum and initial learning rate. \\
Final learning rate & $0.000005$ & $0.000005$ & Minimum and final learning rate. \\
\bottomrule
\end{tabular}
\end{table}

\begin{table}[]
\caption{Hyperparameter values chosen in the set up of Transolver for surface pressure predictions on 2D airfoils in Section \ref{rae2822exp} and Transolver++ on the 3D NASA CRM aircraft configuration in Section \ref{nasacrm}, respectively. Optimized hyperparameters on the 2D airfoil dataset marked in bold.} \label{Tab:tabHPsTransolver}
\begin{tabular}{p{3.5cm} p{2cm} p{2cm} p{7cm}}
\toprule
Hyperparameter & Value (2D) & Value (3D) & Description \\[0.5ex]
\midrule
Hidden size & 128 & 256 & Hidden dimension in the attention mechanism. \\
Heads & 8 & 8 & Number of attention heads. \\
Layers & 8 & 4 & Number of Transolver blocks. \\
MLP ratio & 1 & 2 & Multiplication factor determining the hidden dimension in the MLPs in each Transolver block from the hidden dimension in the attention mechanism. \\
Dropout & 0.0 & 0.1 & Dropout probability in the attention mechanism. \\
\textbf{Slices} & 64 & 64 & Number of slices into which domain points subject to similar physical states are grouped, i.e.\ number of (physics-aware) tokens in the attention mechanism. \\
\bottomrule
\end{tabular}
\end{table}

\begin{table}[]
\caption{Hyperparameter choices in the training procedure of Transolver for surface pressure predictions on 2D airfoils in Section \ref{rae2822exp} and Transolver++ on the 3D NASA CRM aircraft configuration in Section \ref{nasacrm}, respectively. Optimized hyperparameters on the 2D airfoil dataset marked in bold.} \label{Tab:tabHPsTransolverTrain}
\begin{tabular}{p{3.5cm} p{2cm} p{2cm} p{7cm}}
\toprule
Hyperparameter & Value (2D) & Value (3D) & Description \\ [0.5ex]
\midrule
Training epochs & 2,500 & 10,000 & Number of training epochs. \\
\textbf{Batch size} & 8 & 4 & Size of the training batches. \\
Loss function & Rel. L2 & MSE & Relative $L^2$-loss function used in the training process. \\
Optimizer & AdamW & Adam & Optimization algorithm. \\
Weight decay & $0.00001$ & $0.0$ & Weight decay coefficient. \\
Learning rate sched. & OneCycleLR & OneCycleLR & PyTorch OneCycleLR learning rate scheduler. \\
\textbf{Max. learning rate} & $0.0001$ & $0.0001$ & Maximum learning rate achieved after an initial growth period. \\
Final div. factor & $10,000$ & $1,000$ & Division factor determining the final (minimum) learning rate from the maximum learning rate. \\
\bottomrule
\end{tabular}
\end{table}

\begin{table}[]
\caption{Hyperparameter values chosen in the set up of UPT for surface pressure predictions on 2D airfoils in Section \ref{rae2822exp} and the 3D NASA CRM aircraft configuration in Section \ref{nasacrm}, respectively. Optimized hyperparameters on the 2D airfoil dataset marked in bold.} \label{Tab:tabHPsUPT}
\begin{tabular}{p{3.5cm} p{2cm} p{2cm} p{7cm}}
\toprule
Hyperparameter & Value (2D) & Value (3D) & Description \\[0.5ex]
\midrule
Encoder depth & 4 & 4 & Number of transformer blocks in the encoder. \\
Approximator depth & 4 & 4 & Number of transformer blocks in the approximator (processor). \\
Decoder depth & 4 & 4 & Number of transformer blocks in the decoder. \\
Heads & 3 & 3 & Number of attention heads used in all transformer blocks as well as the encoder Perceiver. \\
Latent dimension & 96 & 96 & Feature dimension throughout all latent layers. \\
\textbf{Supernodes} & 36 & 20480 & Number of supernodes to which the input nodes are downsampled. \\
Pooling radius & $0.062$ & $0.142$ & Pooling radius of the supernodes in the encoder GNO. \\
Max. degrees & 32 & 32 & Upper bound for the number of neighbors per supernode. \\
\textbf{Latent tokens} & 36 & 5120 & Number of tokens in the approximator (processor) transformer blocks. \\
\bottomrule
\end{tabular}
\end{table}

\begin{table}[]
\caption{Hyperparameter choices in the training procedure of UPT for surface pressure predictions on 2D airfoils in Section \ref{rae2822exp} and the 3D NASA CRM aircraft configuration in Section \ref{nasacrm}, respectively. Optimized hyperparameters on the 2D dataset marked in bold.} \label{Tab:tabHPsUPTrain}
\begin{tabular}{p{3.5cm} p{2cm} p{2cm} p{7cm}}
\toprule
Hyperparameter & Value (2D) & Value (3D) & Description \\ [0.5ex]
\midrule
\textbf{Training epochs} & 15,000 & 10,000 & Number of training epochs. \\
\textbf{Batch size} & 8 & 1 & Size of the training batches. \\
Loss function & MSE & MSE & Mean squared error loss function used in the training process. \\
Optimizer & AdamW & AdamW & Optimization algorithm. \\
Weight decay & $0.05$ & $0.05$ & Weight decay coefficient. \\
\textbf{Learning rate sched.} & Custom & Custom & Custom Learning rate scheduler comprising two phases of linear growth and exponential decay. \\
Init. learning rate & $0.0$ & $0.0$ & Initial learning rate. \\
\textbf{Max. learning rate} & $0.0002$ & $0.0002$ & Maximum learning rate achieved after the initial warm up epochs (linear growth). \\
\textbf{Decay rate} & $0.0004$ & $0.0004$ & Decay rate of the learning rate during exponential decay phase. \\
\bottomrule
\end{tabular}
\end{table}

\begin{table}[]
\caption{Hyperparameter values chosen in the set up of GAOT for surface pressure predictions on 2D airfoils in Section \ref{rae2822exp} and the 3D NASA CRM aircraft configuration in Section \ref{nasacrm}, respectively. Optimized hyperparameters on the 2D airfoil dataset marked in bold.} \label{Tab:tabHPsGAOT}
\begin{tabular}{p{3.5cm} p{2cm} p{2cm} p{7cm}}
\toprule
\multicolumn{4}{c}{\textbf{Encoder/ Decoder (MAGNO)}} \\[1.0ex]
\midrule
Hyperparameter & Value (2D) & Value (3D) & Description \\[0.5ex]
\midrule
Tokenization strategy & Stencil grid & Stencil grid & The supernodes (tokens) are created as a structured stencil grid in the physical domain. While the authors of \cite{gaot} describe further possible ways of choosing the supernodes, only this method has been implemented in the publicly available code. \\
\textbf{Supernodes} & $[30, 30]$ & $[64, 64, 32]$ & The number of nodes (tokens) in each dimension of the latent stencil grid. \\
Scales & $1$ & $1$ & The number of different pooling radii used in MAGNO. \\
\textbf{Pooling radii} & $0.066$ & $0.066$ & The specific values of the pooling radii used in MAGNO. \\
MLP layers & $3$ & $3$ & Number of layers in each MLP used in MAGNO in both the encoder and decoder. \\
Hidden size & $64$ & $64$ & Hidden dimension in each MLP used in MAGNO in both the encoder and decoder. \\
Lifting channels & $64$ & $64$ & Output dimension of MAGNO. \\
Geometric embedding & statistical & statistical & Geometric information is extracted from neighborhoods of nodes in the form of statistical descriptors. Alternatively, geometric information can be extracted via a PointNet model. \\
\midrule
\multicolumn{4}{c}{\textbf{Processor (transformer)}} \\[1.0ex]
\midrule
Hyperparameter & Value (2D) & Value (3D) & Description \\[0.5ex]
\midrule
Transformer blocks & 3 & 3 &  Number of transformer blocks. \\
Hidden size & $256$ & $256$ & Hidden dimension in the attention mechanism. \\
Patch size & $2$ & $2$ & Patch size for token grouping. \\
Positional embedding & absolute & RoPE & Positional embedding strategy. \\
Heads & 8 & 8 & Number of attention heads. \\
Dropout ratio & 0.0 & 0.2 & Dropout probability in the attention mechanism. \\
Hidden size FFN & 1024 & 1024 & Hidden dimension of the feed forward neural networks in the transformer blocks. \\
\bottomrule
\end{tabular}
\end{table}

\begin{table}[]
\caption{Hyperparameter choices in the training procedure of GAOT for surface pressure predictions on 2D airfoils in Section \ref{rae2822exp} and the 3D NASA CRM aircraft configuration in Section \ref{nasacrm}, respectively. Optimized hyperparameters on the 2D airfoil dataset marked in bold.} \label{Tab:tabHPsGAOTTrain}
\begin{tabular}{p{3.5cm} p{2cm} p{2cm} p{7cm}}
\toprule
Hyperparameter & Value (2D) & Value (3D) & Description \\ [0.5ex]
\midrule
Training epochs & 5,000 & 10,000 & Number of training epochs. \\
\textbf{Batch size} & 8 & 1 & Size of the training batches. \\
Loss function & MSE & MSE & Mean squared error loss function used in the training process. \\
Optimizer & AdamW & AdamW & Optimization algorithm. \\
Weight decay & 0.00001 & 0.00001 & Weight decay coefficient. \\
Learning rate sched. & Mix & Mix & Custom Learning rate scheduler provided by the models' authors comprising three phases of linear growth, cosine decay and exponential decay. \\
Init. learning rate & $0.0008$ & $0.0008$ & Initial learning rate. \\
\textbf{Max. learning rate} & $0.001$ & $0.001$ & Maximum learning rate achieved after the initial warm up epochs (linear growth). \\
Min. learning rate & $0.0001$ & $0.0001$ & Learning rate achieved after a dropdown from the max. learning rate following a cosine decay. \\
Final learning rate & $0.00005$ & $0.00005$ & Learning rate achieved after a dropdown from the min. learning rate following a final exponential decay. \\
\bottomrule
\end{tabular}
\end{table}

\clearpage

\end{document}